\pdfoutput=1

\documentclass[11pt]{article} 

\usepackage{natbib}

\usepackage[T1]{fontenc}
\usepackage{fancyhdr}
\usepackage{amsmath}
\usepackage{amssymb}
\usepackage{amsthm}
\usepackage{enumitem}
\usepackage{epsfig}
\usepackage{graphicx}
\usepackage{url}
\usepackage{exscale}

\usepackage{color}

\usepackage{booktabs} 

\newtheorem{defn}{Definition}

\newcommand{\mean}{{\mathbf E}}

\newcommand{\note}[1]
{$^{(!)}$\marginpar[{\hfill\tiny{\sf{#1}}}]{\tiny{\sf{(!) #1}}}}

 \def\doublespace{\baselineskip=\normalbaselineskip \multiply\baselineskip by
 3 \divide\baselineskip by 2}

\begin{document}

\doublespace

\title{On the relationship between the theory of cointegration and the theory of phase synchronization
\footnotetext{Address for correspondence: Institute of Applied Mathematics, University
of Heidelberg, Im Neuenheimer Feld 294, D-69120 Heidelberg, Germany (e-mail: dahlhaus@statlab.uni-heidelberg.de).} }

\author{Rainer Dahlhaus, Istv\'{a}n Z. Kiss, and Jan C. Neddermeyer\\ \normalsize{Heidelberg University, Saint Louis University, and DZ BANK AG} \\ }

\date{May 11, 2018}

\begin{titlepage}
\maketitle
\thispagestyle{empty}

\noindent
{\small \textbf{Summary.}
The theory of cointegration has been a leading theory in econometrics with powerful applications to macroeconomics during the last decades. On the other hand the theory of phase synchronization for weakly coupled complex oscillators has been one of the leading theories in physics for many years with many applications to different areas of science.
For example, in neuroscience phase synchronization is regarded as essential for functional coupling of different brain regions. In an abstract sense both theories describe the dynamic fluctuation around some equilibrium. In this paper, we point out that there exists a very close connection between both theories. Apart from phase jumps, a stochastic version of the Kuramoto equations can be approximated by a cointegrated system of difference equations. As one consequence, the rich theory on statistical inference for cointegrated systems can immediately be applied for statistical inference on phase synchronization based on empirical data. This includes tests for
phase synchronization, tests for unidirectional coupling and the identification of the equilibrium from data including phase shifts. We study two examples on a unidirectionally coupled R\"{o}ssler-Lorenz system and on electrochemical oscillators. The methods from cointegration may also be used to investigate phase synchronization in complex networks. Conversely, there are many interesting results on phase synchronization which may inspire new research on cointegration.}

\bigskip


\end{titlepage}

\tableofcontents

\setcounter{page}{0}

\newpage

\section{Introduction}

Phase Synchronization has a long history dating back to 1665, where the mathematician and
physicist C. Huygens discovered synchronization of two pendulum
clocks suspended close to each other on the same wooden beam. From that time on, the phenomenon got
more and more into the focus of scientists. During the last decades the behavior of two or several
interacting oscillators has been intensively studied in the physics literature - in particular in the context of nonlinear dynamical systems (cf. \cite{Pec-1990,Koc-1996,Ros-1996,Boc-2001,Boc-2002};  - see also the monographs \cite{Pik-2001,Osi-2007}. In contrast to other types of synchronization, phase synchronization purely depends on the phases of self-sustained weakly-coupled oscillators while the amplitudes may even be empirically uncorrelated. Weak coupling means that the phases of all oscillators may be subject to individual disturbances and the whole system adjusts itself afterwards. Thus phase synchronization is  regarded as a complex dynamical process rather than a fixed state.

The phenomena of phase synchronization has been found experimentally in many fields, e.g. electrical circuits in \cite{Puj-2003,Bap-2003}, lasers in \cite{DeS-2001}, electrochemistry in \cite{Kis-2001,Kis-2002}, biological systems in \cite{Els-1998,Tas-1998}, population dynamics in \cite{Bla-1999}, between El Ni\~{n}o and the Indian monsoon in \cite{Mar-2005} and even tennis in \cite{Pal-2005}. Since phase synchronization purely depends on the phases of the oscillators it can also be detected between oscillators which are of a different type. A practical example  is the phase synchronization of the cardiac and respiratory system in \cite{Lot-2000,Ste-2000} or between maternal breathing and the fetal-maternal heart rate coordination in \cite{VanLee-2009}.

In neuroscience, phase synchronization is regarded as essential for functional coupling of different brain regions. Single neuronal activity,  coupled to an ensemble of oscillatory neuronal activity through the phase, enables the cells to transmit their information content long-range across different cortical areas. To cite from the abstract of  \cite{Fel-2011}: ``In recent years, studies ranging from single-unit recordings in animals to EEG and MEG studies in humans have demonstrated the pivotal role of phase synchronization in memory processes. Phase synchronization - here referring to the synchronization of oscillatory phases between different brain regions - supports both working memory and long-term memory and acts by facilitating neural communication and by promoting neural plasticity.'' \cite{Col-2010} discuss mechanisms of gamma oscillations in the hippocampus and the functional role of the synchronization of such oscillations in several key hippocampal operations, including cell grouping, dynamic routing, and memory. In \cite{Wom-2007} synchronization in the gamma band is investigated, and it is discovered that the mutual influence among neuronal groups depends on the phase relation between rhythmic activities within the groups. Furthermore, the pattern of synchronization is related to the pattern of neuronal interactions. Further reviews on phase synchronization in neuroscience are \cite{Eng-2001,Var-2001,Dav-2003}.

To study the synchronization of a larger population of interacting units theoretically - for example fireflies flashing or crickets chirping at the same time -  \cite{Win-1967} studied the nonlinear dynamics of a large population of weakly coupled oscillators with intrinsic frequencies that were distributed according to some prescribed probability distribution. He also ignored the amplitude and considered phase oscillators and worked with a mean field model.  \cite{Kur-1975,Kur-1984} introduced a sound mathematical model to describe this phenomenology leading to a theoretical treatment of the mean field approach. He also studied the limit behavior when the number of oscillators tends to infinity. This model has been used since then in different forms to discuss theoretically phase synchronization in a population of weakly coupled oscillators. For example the onset of synchronization is discussed in this framework in \cite{Str-2000}.

Although some stochastic methods are used to analyze phase synchronization (e.g. the spectral coherence), there hardly exist any stochastic models for the dynamics of synchronized phase processes. Establishment of experiment-adapted stochastic phase models would be an important task to estimate the dynamics of the phases. Specific outcomes could be statistical tests for
phase synchronization, the presence of unidirectional coupling, and the identification of the equilibrium from data including phase shifts.

In this paper, we propose that the theory of cointegration provides a good stochastic model for phase synchronization and therefore is a good framework for investigating these questions. In the method, cointegration is not used directly
as a model for the oscillators, but rather as a model for the phase processes driving the oscillators. Using the phase processes as the key element for describing oscillators has been standard in physics for many years. On the other hand, this approach has never been used to our knowledge in statistics or econometrics to model oscillators. Here oscillators have typically been investigated with spectral methods such as Fourier or wavelet transforms. Furthermore, statisticians focused on building models directly for the oscillating processes instead of the phases.

The theory of cointegration may be used in phase synchronization both for theoretical studies and for identifying unknown systems based on empirical data. The link may also be of importance in the other direction in that the known results on phase synchronization may lead to new insights or stimulate research in cointegration. The common ground of both theories is that two or several processes fluctuate randomly around some equilibrium.

The concept of cointegration was introduced by \cite{Gra-1981} for the joint modeling of several macroeconomic variables over time. In 2003 he received the Nobel Prize for his discovery jointly with  Engle. The aim of the theory is to model common stochastic trends, for example of income and consumption where the short-run dynamics is affected by random disturbances and the long-run dynamics is restricted by economic equilibrium relationships. Other examples are the exchange rates and the price levels, the short and long-term interest rates, and the
prices on spot and future markets.

Since its introduction the theory of cointegration has been developed by many researchers and it has become a leading theory in econometrics with many applications in macroeconomics and beyond - for references see the monographs \cite{Ban-1993,Joh-1995,Eng-1999,Lue-2005,Jus-2006} among many others. \cite{Phi-1991,Kes-2001} have developed the theory for continuous time diffusion processes.


We now give a heuristic and elementary argument why the setting of cointegration is of importance for phase synchronization:
Suppose we observe  $y_{t}^{(i)} = A_i \cos (\phi_{t}^{(i)}) + \varepsilon^{(i)}_t$ for $i=1,2$ and $t=1,\ldots,T$ with phase processes $\phi_{t}^{(i)}$
(alternatively we may just have the phases $\phi_{t}^{(i)}$ from some oscillators calculated by means of the Hilbert transform or some other method).
A naive model for a random phase could be $\phi_t = \omega t + \phi_0 + \delta_t$ with some random error $\delta_t$ which could be a white noise process
or some stationary process. This model had the drawback that it fluctuates around the deterministic linear phase $\omega t + \phi_0$ and would therefore
only be adequate if some external force or constraint would keep the phase close to this linear phase. Instead a better model for most situations is the
corresponding model for the phase increments $\Delta \phi_t =  \phi_t - \phi_{t-1} = \omega  + \delta_t$ where $\delta_t$ is a stationary process with
mean zero and positive correlation (in a more refined model one would in addition request positivity of the phase increments - cf. Appendix~\ref{sec:Appendix1}). In the simplest case where $\delta_t$ is iid Gaussian this means that $\phi_t = \omega t + \phi_0 + \sum_{s=1}^{t} \delta_s$
has the same distribution as a Brownian motion with drift. In this case ${\rm var} (\phi_t) \sim t$ which means that the phase may depart substantially from
$\phi_t = \omega t + \phi_0$ for large $t$. This feature remains if the process $\delta_t$ is stationary with mean $0$. Such a process $\phi_t$ is called an integrated process.

If we look at two processes with synchronized phases then both phases may evolve like a Brownian motion with drift - however the difference
$\phi_{t}^{(1)} - \phi_{t}^{(2)}$ should stay relatively small; in particular it should not explode. A proper model therefore is that these differences
follow a stationary process. This is exactly the concept of cointegration: both phase-processes are integrated but the difference remains stationary.
The heuristics presented here is formalized in Appendix~1.

The paper is organized as follows. Section~\ref{sec:physics} provides a brief
introduction to phase synchronization with focus on testing for phase synchronization and the Kuramoto model for interacting oscillators. We point out that a cointegrated system arises as the solution of a system of stochastic difference equations which is similar to the Kuramoto equations. In Section~\ref{sec:cointegration}, we give a brief introduction to the theory of cointegration and review some tests for cointegration.
Section~\ref{sec:newdefinition} describes the use of cointegration for the statistical analysis of phase synchronization. In particular we give our definition of stochastic phase synchronization.  In Section~\ref{sec:examples} we present a simulation showing the application of the
proposed method to a coupled R\"{o}ssler-Lorenz system, and an experimental example of electrochemical
oscillators. Conclusions and an outlook are given
in Section~\ref{sec:conclusion} followed by the definition of VEC-state oscillators in Appendix~1 and some computational and modeling aspects in Appendix~2.

%
%
%


\section{Phase synchronization and the Kuramoto model for interacting oscillators}\label{sec:physics}

\subsection{Phase synchronization of weakly coupled oscillators}\label{sec:phasesynchronization}

The starting point of phase synchronization is the definition of the phase $\phi_t$ of an oscillator. For chaotic phase synchronization, the phase often is defined via the Hilbert transform of the signal. If the projection of the attractor on some plane has only one rotation center (as in Figure~\ref{fig:RoesslerLorenzProjections}) then the phase can be defined by the  rotation angle around this center. The
Hilbert phase of a signal $y_t$ is defined through the analytic signal
$\zeta_t$ given by $\zeta_t = y_t + i y_t^H = a_t \exp(i \phi_t)$. The imaginary part of $\zeta_t$ is obtained using the Hilbert transform - for a definition in the continuous case c.f. \cite{Pik-1997}, Appendix A.2, and for a definition in the discrete time case  c.f. \cite{Brill-2001}, Section  2.7. $a_t$ is the amplitude, and the phase $\phi_t$ is computed through
\[
\phi_t = \text{arctan}
\frac{y_t^H}{y_t}.
\]
Alternatively, one may look at stochastic systems, e.g. of the form
\begin{equation} \label{CosineOscillator}
y_t = a_t \cos(\phi_t) + \varepsilon_t, \quad t \in \mathbb{Z}
\end{equation}
or even at systems like
\begin{equation}\label{GeneralOscillator}
y_t = a_t f(\phi_t) + \varepsilon_t, \quad t \in \mathbb{Z}
\end{equation}
where $f$ is a $2 \pi$ periodic real valued function representing the oscillation pattern. Here the $\phi_t$ may be determined by a particle filter (\cite{Dah-2017}). The latter model may for example be used for ECG signals. Further methods for phase estimation are based on the wavelet transform \cite{Gro-1989} or on a local periodogram (\cite{Han-1973},\cite{Para-2012}).

In the next step two weakly coupled oscillators with phases $\phi_t^{(1)}$ and $\phi_t^{(2)}$ are called $m:n$ phase synchronized if there exists some constant $\Phi$ (phase shift) and some small $\delta>0$ such that
\begin{equation}\label{weaklockingcondition}
\big| \big(m \phi_t^{(1)} - n\phi_t^{(2)} - \Phi \big) {\rm mod} \, 2\pi \big| < \delta
\end{equation}
holds for all $t$. $m$ and $n$ are integers which are
usually known in a practical application. The idea behind this definition is that the (rescaled) phases ``stay together'' and do not move arbitrarily far from each other. A standard statistic for testing  the hypothesis ``no phase synchronization'' is the phase synchronization index \cite{Mor-2000,Qui-2002} defined through
\begin{equation} \label{PhaseSynchroIndex}
\hat{R}^{2} = \bigg| \frac {1} {T} \sum_{t=1}^{T} \exp \big\{ i \big( m \phi_{t}^{(1)} - n \phi_{t}^{(2)}\big)\big\} \bigg|^{2}.
\end{equation}
The synchronization index has values between $0$ and $1$ where values close to one indicate phase synchronization. A value close to 1 is obtained for an almost constant phase difference of $\big|m \phi_{t}^{(1)} - n \phi_{t}^{(2)}\big|$, which can occur even for two nonidentical chaotic oscillators (\cite{Ros-1996}). $\hat{R}^{2}$ has often been used for testing phase synchronization and there exist several articles where the distribution of $\hat{R}^{2}$ or other test statistics is approximated by different methods - e.g. with surrogate data. The most advanced approach from a statistical perspective is \cite{Schel-2007} where the distribution of $\hat{R}^{2}$ under the null hypothesis is approximated by using two independent stochastic processes. Reviews and comparisons of various phase synchronization measures and tests can be found in \cite{Qui-2002},  \cite{All-2004a}, and \cite{Schel-2007}.

Despite of these results, the situation is not satisfying from a statistical perspective, because there
usually does not exist a clearly defined population quantity corresponding to $\hat{R}^{2}$. We show in this paper that cointegrated systems provide a statistical framework with which questions like existence of synchronization and unidirectional coupling can be investigated.


\subsection{Cointegration as a stochastic Kuramoto-type model}\label{sec:Kuramoto}

As mentioned in the introduction  \cite{Kur-1975,Kur-1984} introduced a mathematical model to describe  the synchronization of a larger population of interacting units. A simple version of the Kuramoto model for the phases of $d$ oscillators is
\begin{equation} \label{Kuramoto}
\dot{\phi}^{(i)} = \omega_i + \frac {K} {d} \sum _{j=1}^{d} \sin \big(\phi^{(j)} - \phi^{(i)}\big) \qquad (i=1,\ldots,d).
\end{equation}
\cite{All-2004b} discuss the following stochastic generalization of the model
\begin{equation} \label{stochasticKuramoto}
\dot{\phi}^{(i)} = \omega_i +  \sum _{j=1}^{d} k_{ij}\, \sin \big(\phi^{(j)} - \phi^{(i)}\big) + \eta_i \qquad (i=1,\ldots,d)
\end{equation}
where the $\eta_i$ are taken to be independent Gaussian white noise variables.

There is a striking similarity of this model to cointegration: For $\phi^{(j)} \!-\! \phi^{(i)}$ small we can approximate $\sin (\phi^{(j)} \!-\! \phi^{(i)})$ by $\phi^{(j)} \!-\! \phi^{(i)}$ leading to the discretized approximation for the vector $\phi_t = \big(\phi^{(1)}_t,\ldots,\phi^{(d)}_t\big)'$, $\omega = \big(\omega_1,\ldots,\omega_d\big)'$
\begin{equation} \label{KuramotoCointegration}
\Delta \phi_t = \omega +  \Pi \, \phi_{t-1} + \eta_t
\end{equation}
with $\Pi=K \!-\! {\rm diag}\big(\sum_{j=1}^{d} \!k_{1j},\ldots,\sum_{j=1}^{d} \!k_{dj}\big)$ where $K = (k_{ij})_{i,j=1,\ldots,d}$. This is exactly the representation (\ref{Pi-representation}) from below with $p=1$  and $\omega=\mu$, i.e. the cointegrated system (\ref{MA-representation}),  (\ref{VEC-representation}) with $p=1$ is exactly the solution of the stochastic difference equations (\ref{KuramotoCointegration}). We mention that cointegration requires the matrix $\Pi$ to be of reduced rank which is fulfilled in this case since the columns sum up to $0$.

Being a bit more specific at this point gives us a deeper insight into the relevance of cointegration theory for phase synchronization in networks: The above $\Pi$ can be written in the form $\Pi = \alpha \beta'$ with $d \times r$ - matrices $\alpha$ and $\beta$ of full rank $r$ where (in this case) $r=d-1$, namely
\begin{equation*} \label{}
\beta' =
\left(
\begin{matrix}
\beta_1'\\
\vdots\\
\beta_r'
\end{matrix}
\right) =
\left(
\begin{matrix}
1 & 0 & \cdots & 0 & -1 \\
0 & 1 & \cdots & 0 & -1 \\
\vdots & & \ddots & & \vdots \\
0 & 0 & \cdots & 1 & -1
\end{matrix}
\right) \qquad \quad
\end{equation*}
and $\alpha$ consisting of the first $d-1$ columns of $\Pi$ (only for this specific $\beta$). The $\beta_j' \phi_t = 0$ are obviously the $d-1$ phase synchronization relations $\phi^{(1)}_t=\cdots=\phi^{(d)}_t$. The matrix $\alpha$ is sometimes called adjustment or loading matrix.

In the following we apply the concept of cointegration to phase synchronization. For the model (\ref{KuramotoCointegration}) this means testing for cointegration (phase synchronization) via determining (testing) $r= {\rm rank} \, \Pi$. This results in the reduced rank representation $\Pi \!=\! \alpha \beta'$ with $d \times r$ - matrices $\alpha$ and $\beta$ of full rank $r$. After the parameters $\alpha$ and $\beta$ have been estimated $\beta' \phi_t$ are the $r$ cointegration (phase synchronization) relations and $\alpha$ gives the intensities with which deviations from the equilibrium lead to corrections. In particular $\alpha$ can be tested for unidirectional coupling.




The use of the cointegration model (\ref{KuramotoCointegration}) for phase synchronization has several benefits: The  model can be fitted in both situations (phase synchronization  / non phase synchronization ) to the data and the hypothesis of phase synchronization can be tested on the matrix $\Pi$ - e.g. by looking at the fitted model. In addition one is able to identify the parameters from real data, and to conclude to the phase synchronization relations, unidirectional coupling etc. This can easily be done by using the large number of existing tools for cointegration.

On the other hand, it is very important to keep the difference between the two models in mind: The major difference is that the Kuramoto model (\ref{stochasticKuramoto}) has the equilibrium $\phi^{(j)} - \phi^{(i)} \equiv 0 \mbox{ mod } 2\pi$ while the cointegration model (\ref{KuramotoCointegration}) has the equilibrium $\phi^{(j)} - \phi^{(i)} = 0$. In case of (say) $\pi/2 < \phi^{(j)} - \phi^{(i)} < \pi$ the Kuramoto model has a force to the ``equilibrium'' $\phi^{(j)} - \phi^{(i)} = 2 \pi$ while the cointegration model still has a force to $\phi^{(j)} - \phi^{(i)} = 0$. This means that the Kuramoto model allows for phase jumps of one of the two processes while the cointegration model does not. One may term the former ``weak synchronization'' and the latter ``strong synchronization''. Examples are the processes with $\epsilon = 9.6, 10.2, 11.0, 11.1$ in Section~\ref{sec:application} (Figures~\ref{fig:ComparisonToPSIndex} and Figure~\ref{fig:RoesslerLorenzPhaseDiff}) which exhibit phase jumps.


\section{The concept of cointegration}\label{sec:cointegration}

\subsection{Cointegration and the Granger representation theorem} \label{sec:cointegrationGranger}

We now briefly review a small part of the theory of cointegration (for more details see e.g the monographs \cite{Joh-1995,Lue-2005,Jus-2006} - the last one containing a more applied view). As it is usual in many papers we restrict ourselves to vector autoregressive (VAR-) processes
\begin{equation} \label{VAR-representation}
X_t = A_1 X_{t-1} + \ldots + A_k X_{t-p} + \mu + \eta_t
\end{equation}
where the coefficients $A_i$ are $d \times d$-matrices, $\mu = (\mu_1,\ldots,\mu_d)'$ and the innovations $\eta_t$ are iid with mean $0$ and positive definite covariance matrix $\Omega_{\eta}$. Let $A (z) := I_{d}-A_1 z - \cdots - A_{p} z^{p}$ be the characteristic polynomial. We restrict ourselves to VAR-processes with roots outside the unit circle or equal to $1$, i.e. where $\det \big(A(z)\big)=0$ implies $|z| \!>\! 1$ or $z=1$.

We  call a process integrated (of order $1$ - $I(1)$ for short) if $X_t$ is not stationary and the first order difference $\Delta X_t = X_t - X_{t-1}$ is stationary. We call a $d$-dimensional process $X_t$ cointegrated if each univariate series is integrated and some linear combination $\beta' X_t$ (with an $r \times d$-matrix
$\beta$ of rank $r$ and $0\!<\!r\!<\!d$) is stationary  (these definitions simplify the issue -
for thorough definitions we refer to the above monographs).
$\beta$ is known as the cointegrating vector. It specifies the long-term relationship between the involved
univariate series. The most relevant example for phase synchronization of $2$ oscillators is $d=2$ and $\beta =(1,-1)'$: $\beta' X_t$ fluctuates stationary around a long run mean - i.e. it is bounded in probability.

With some straightforward calculations we can transform the above representation to
\begin{equation} \label{Pi-representation}
\Delta X_t = \Pi \, X_{t-1} + \sum_{i=1}^{p-1} \Gamma_i \,
\Delta X_{t-i} + \mu + \eta_t
\end{equation}
where $\Pi := -(I_d - A_1 - \cdots - A_p)$ and $\Gamma_i := - (A_{i+1}+\cdots+A_{p})$ for $i=1,\ldots,p-1$.
If the process $X_{t}$ is not stationary (in particular if it is cointegrated) then $\det (\Pi) = (-1)^{d} \det \big(A(1)\big)=0$. Thus $\Pi$ is singular with rank $\,r \!<\! d$ and it can be decomposed to $\Pi = \alpha \beta'$ with $d \times r$ - matrices $\alpha$ and $\beta$ of full rank $r$.

Since the process is $I(1)$, the first order difference $\Delta X_t$ is stationary, and the representation (\ref{Pi-representation}) implies that also $\Pi X_{t-1} = \alpha \beta' X_{t-1}$ is stationary. Multiplying this with $(\alpha'\alpha)^{-1} \alpha'$ implies that $\beta' X_{t-1}$ is stationary - i.e. each component of $\beta' X_{t-1}$ is a cointegrating relation. The corresponding representation
\begin{equation} \label{VEC-representation}
\Delta X_t = \alpha \, \beta'  X_{t-1} + \sum_{i=1}^{p-1} \Gamma_i \,
\Delta X_{t-i} + \mu + \eta_t.
\end{equation}
is called \textit{vector error correction model} (VEC-model or VEC-representation). It shows that whenever the process moves away from the equilibrium $\beta'  X_{t-1} = \beta_0$ (for the definition of $\beta_0$ see below) it is pulled back to the equilibrium with the forces $\alpha$ (see the end of Section~\ref{sec:application} for an illustrative example).

The Granger representation theorem (\cite{Eng-1987,Joh-1991}) now gives another useful representation
of the process. Under the above conditions the process has the moving average representation
\begin{equation} \label{MA-representation}
X_t = C \sum_{i=1}^{t} \eta_i + C \mu t + C^{*}(L) (\eta_t
 + \mu) + X_{0}^{*}
 \end{equation}
where
\begin{equation} \label{Definition C}
C = \beta_{\bot} \Big[\alpha_{\bot}' \Gamma \beta_{\bot}\Big]^{-1} \alpha_{\bot}'
\end{equation}
with $\Gamma := I_{p}-\sum_{i=1}^{p-1} \Gamma_{i}$. $C^{*}(L) (\eta_t + \mu) = \sum_{j=0}^{\infty} C_{j}^{*} (\eta_{t-j} + \mu)$ is a stationary process and $X_{0}^{*}$ contains initial values with $\beta' X_{0}^{*}=0$. $\alpha_{\bot}$ denotes an orthogonal complement of $\alpha$ i.e. $\alpha_{\bot}$ is an $d \times (d-r)$-matrix of rank $d-r$ with $\alpha_{\bot}' \alpha =0$ (the same for $\beta_{\bot}$).

The representation (\ref{MA-representation}) shows that the $d$-dimensional process is driven by $d\!-\!r\,$ $I(1)$ components and $r$ stationary components. The first term on the right hand side consists of $d$ random walks $\sum_{i=1}^{t} \eta_i$ which are multiplied by a matrix of rank $d-r$ denoted by $C$. Thus, there are actually $d-r$ stochastic trends driving the system. On the contrary the representation (\ref{VEC-representation}) (or (\ref{VEC-representation 2}) below) shows how the process is pulled back to the equilibrium if deviations occur. This is illustrated by the example at the end of Section~\ref{sec:application}.

Of special importance for phase synchronization is a decomposition $\mu = \Gamma \omega_0 - \alpha \beta_0$ where
$\omega_0$ is a trend-term and $\beta_0$ is a constant belonging to the error correction equation:
Taking expectations in (\ref{MA-representation}) we obtain because of $\mean \eta_t = 0$
\begin{equation} \label{Mean-MA-representation}
\mean X_t = C \mu t + C^{*}(L) \, \mu + \mean X_{0}^{*}
\end{equation}
and therefore $\omega_0 := \mean \Delta X_t = C \mu$. Since $\Delta X_t$ is stationary we obtain from (\ref{VEC-representation})
\begin{equation*} \label{}
\alpha \, \mean \big(\beta'  X_{t-1} \big) = \mean \big(\Delta X_t \big) - \sum_{i=1}^{p-1} \Gamma_i \,
\mean \big(\Delta X_{t-i}\big) - \mu = \big[\Gamma C - I_d \big] \mu.
\end{equation*}
Let $\bar{\alpha} := \alpha (\alpha' \alpha)^{-1}$. Since $\bar{\alpha}'\, \alpha = I_r$ we have
$\mean \big(\beta'  X_{t-1} \big) = \bar{\alpha}' \big[ \Gamma C - I_d \big] \mu =: \beta_0$.

Obviously we have $\mu = \Gamma C \mu - \big[ \Gamma C - I_d \big] \mu = \Gamma \omega_0 - \alpha \beta_0$.
Therefore we obtain the modified VEC-representation%
\begin{equation} \label{VEC-representation 2}
(\Delta X_t - \omega_0) = \alpha \, \big(\beta'  X_{t-1} - \beta_0\big) + \sum_{i=1}^{p-1} \Gamma_i \,
\big(\Delta X_{t-i} - \omega_0 \big) + \eta_{t}.
\end{equation}
This means that the ``true'' cointegration relation describing the equilibrium is $\beta'  X_{t-1} - \beta_0 = 0$. If the process deviates at time $t-1$ from this relation it is pulled in the next step with force $\alpha$ back towards this equilibrium. (\ref{MA-representation}) and (\ref{Mean-MA-representation}) show that $\omega_0 = C \mu$ is the drift-vector of the process, that is the process $X_t$ has a deterministic trend $\omega_0 t$. The intercept terms are contained in $X_{0}^{*}$ in (\ref{MA-representation}).

Below we use this model as a stochastic model for phase synchronization. In the case $d=2$ the situation is even more intuitive: If the system is cointegrated (i.e. $r=1$) we only have a one-dimensional cointegration relation $\beta'  X_{t-1} - \beta_0 = 0$ and the drift vector $\omega_0$ has the same direction as the random walk part $C \sum_{i=1}^{t} \eta_i$ (namely $\beta_{\bot}$). We discuss this simpler and more intuitive case in Section~\ref{sec:application} with a specific example.

To keep the situation simple we first restrict ourselves to the case $p=2$. In this case the VEC-representation takes the form (now replacing $X_t$ by $\phi_t$)
\begin{eqnarray}
\label{vecm1}
\Delta \phi_t^{(1)} &=& \alpha_1 \big( \phi_{t-1}^{(1)} - \beta_2 \phi_{t-1}^{(2)}\big) +
\gamma_1 \Delta \phi_{t-1}^{(1)} + \delta_1 \Delta \phi_{t-1}^{(2)} + \mu_1 +
\eta^{(1)}_t,\\
\label{vecm2}
\Delta \phi_t^{(2)} &=& \alpha_2 \big(\phi_{t-1}^{(1)} - \beta_2 \phi_{t-1}^{(2)}\big)
 + \gamma_2 \Delta \phi_{t-1}^{(2)} + \delta_2 \Delta \phi_{t-1}^{(1)} + \mu_2 +
\eta^{(2)}_t,
\end{eqnarray}
where we have assumed for the cointegrating vector $\beta = (1, -\beta_2)'$ for identifiability. In that case also $\alpha_1$ and $\alpha_2$ are identifiable.

If the model is used for phase synchronization $\beta_2$ usually is known (in the $1\!:\!1$ synchronization case we have $\beta_2\!=\!1$; $\;\,\beta_0$ then is the mean phase difference in the equilibrium). We expect that in many cases the reduced system with $\delta_1=\delta_2=0$ will suffice to describe the joint phase dynamics. However, the parameters $\gamma_1, \gamma_2$ are needed since successive increments usually are correlated.

In the reduced system where $\delta_1=\delta_2=0$ error correction towards the equilibrium is done solely by the terms with coefficients $\alpha_1$ and $\alpha_2$. Of special importance are the cases $\alpha_1 < 0, \alpha_2 = 0$ and $\alpha_1 = 0, \alpha_2 > 0$ meaning that there is unidirectional coupling - see Section~\ref{sec:application} for an illustrative example
(in the non-reduced system also the higher order terms contribute to error correction - cf.  \cite{Joh-1995}, Exercise 4.3). This allows for testing whether there is an unidirectional driver-response relationship between the phases.

A general state space model is discussed in Appendix~\ref{sec:Appendix1}. There also the positivity of the phase increments is discussed.


\subsection{Cointegration Testing and Model Fitting}\label{sec:cointegrationtests}

In order to test for phase synchronization and for unidirectional driver-response relationships we apply tests for cointegration. Two cases need to be distinguished:
First the case where the cointegrating vector $\beta$ is known and we want to test for a specific cointegrating relationship. This may be the standard case for phase synchronization of two oscillators where the $m\! :\! n$ synchronization relation as in (\ref{weaklockingcondition}) is often clear from prior knowledge or eye-inspection.

The other case is the case of unknown $\beta$. For example in systems of higher dimension it may be clear that the whole system is ``somehow'' $1\!:\!1$ synchronized but it usually is not clear at all how the synchronization ``propagates'' through the network. In that case the task is to test for the rank of $\Pi$ in (\ref{Pi-representation}) and to determine the factorization $\Pi = \alpha \, \beta'$ as well as the vector $\beta_0$ leading to the phase synchronization relations, the mean phase shift and the matrix $\alpha$ representing the strength and the direction of coupling.

We start with the case of unknown $\beta$ leading to the likelihood theory of \cite{Joh-1995} and to his likelihood ratio test. Additional information on this case is provided in Section~\ref{sec:newdefinition}. We then discuss the augmented unit root test by \cite{Dic-1979,Dic-1981}. We mention that there exist several other tests such as a test based on residuals by \cite{Phi-1990} and Lagrange multiplier tests (\cite{Sai-2000}).

\subsubsection*{Johansen's likelihood theory and rank test}\label{sec:JohansenRankTest}

The likelihood theory for cointegrated systems is well developed - the most famous result being Johansen's likelihood ratio test (LR-test) for the cointegration rank - i.e for the determination of $r = {\rm rank}(\Pi)$ in (\ref{Pi-representation}) and the corresponding
decomposition $\Pi = \alpha \, \beta'$ with $\alpha$ and $\beta$ being $d \times r$ - matrices of rank $r$. Thus a major advantage of the Johansen procedure is that it can simultaneously identify multiple cointegration relations in multivariate time series. The situation is much more challenging than for the classical likelihood ratio test since the integrated processes require a different asymptotic theory leading for example to a nonstandard limit distribution of the likelihood ratio test.

We briefly sketch part of the results - for a more detailed discussion cf. Chapters 7 and 8 in  \cite{Jus-2006}. Note that in our case $\mu$ in (\ref{Pi-representation}),(\ref{VEC-representation})  plays an important role since it leads to the phase lag $\beta_0$ and the deterministic drift term $\omega_0 t$ of the phases, while the additional term  $\mu_1 t$ often discussed in the cointegration literature (leading to quadratic drift) is not needed in this context (i.e. the situation we consider is \textit{Case 3} in \cite{Jus-2006}).

Suppose now we want to calculate the maximum likelihood estimates for $\alpha$ and $\beta$ in the system (\ref{VEC-representation}). In the Gaussian case the maximum likelihood estimates are essentially the same as least squares estimates. In the first step  $\mu$ and the $\Delta X_{t-i}$ are removed by regressing them on $\Delta X_t$ and  $X_{t-1}$. If we denote the residuals by $\widetilde{\Delta X_t}$ and  $\widetilde{X_{t-1}}$ we obtain the ``concentrated model''
\begin{equation*} \label{}
\widetilde{\Delta X_t} = \alpha \, \beta' \widetilde{X_{t-1}} + \mbox{error}.
\end{equation*}
Now, the optimal value $\hat{\alpha}(\beta)$ is determined for given $\beta$, plugged into this equation, and finally the optimal $\beta$ is determined leading to the maximum of the likelihood. This procedure can be viewed as finding the canonical correlations between $\widetilde{\Delta X_t}$ and $\widetilde{X_{t-1}}$ (c.f. \cite{Jus-2006}, Chapter 8). The resulting squared canonical correlations are denoted by $\hat{\lambda}_1 \ge \ldots \ge \hat{\lambda}_d \ge 0$. The task now is to split the $\hat{\lambda}_1 \ge \ldots \ge \hat{\lambda}_d$ into those $\hat{\lambda}_1 , \ldots , \hat{\lambda}_r$ which are different from zero and which belong to a cointegration relation and those $\hat{\lambda}_{r+1} , \ldots ,\hat{\lambda}_d$ which are not significantly different from zero (more precisely where the corresponding unknown theoretical values are zero). With view on (\ref{MA-representation}) and the discussion below, $d-r$ is the number of driving ``forces'' of the trend. In particular $r$ is the rank of the matrix $\Pi$ we are looking for.

A formal way to determine $r={\rm rank}(\Pi)$ based on the values of $\hat{\lambda}_1,\ldots , \hat{\lambda}_d$ is the likelihood ratio test - e.g. the so-called trace-test where the null hypothesis $\mathcal{H}(r) = \{{\rm rank}(\Pi) \le r\}$ is tested in $\mathcal{H}(d):= \{{\rm rank}(\Pi) \le d\}$  recursively for $r=0,\ldots,d-1$ (this recursion is called ``top $\rightarrow$ bottom'' procedure with ``top'' meaning a large number of cointegration relations corresponding to a small rank). The test statistic is
\begin{equation*} \label{}
\tau_{d-r} := LR \big(\mathcal{H}(r) | \mathcal{H}(d)\big) = -T \sum_{i=r+1}^{d} \log (1-\hat{\lambda}_i)
\end{equation*}
(called LR-test below) and the hypothesis is rejected at a significance level $\alpha$ if $\tau_{d-r} > C_{\alpha}(d-r)$ where $C_{\alpha}(d-r)$ is from Appendix A of \cite{Jus-2006} (Case 3; $p\!-\!r=d\!-\!r$). The procedure stops if the test accepts the hypothesis for some $r$.

Special attention is required for the case $r=d$. This means that there are no stochastic trends and the system in (\ref{Pi-representation}) is stationary. In particular this case does not mean that there are $r$ cointegration relations - see the discussion in the next section.

\subsubsection*{Augmented Dickey-Fuller Unit Root Test}\label{sec:adf}

The Augmented Dickey-Fuller (ADF) test allows one to test for the presence of a
unit root in a univariate time series $Y_t$. It is based on the regression
\begin{equation*} \label{ADF-regression}
\Delta Y_t = a + bt + cy_{t-1} + \sum_{j=1}^p c_j \Delta Y_{t-j} + \epsilon_t.
\end{equation*}
The constant $a$ and the time trend $bt$ are only included if required.
For $a \neq 0$ and $b=0$ the ADF-test tests the null hypothesis $c=0$ \textit{integrated process with a deterministic trend} against the alternative $c\!<\!0$ \textit{stationarity with a mean}. The test statistic is the usual t-ratio given by
\[
\text{DF}_{\tau} = \frac{\hat{c}}{\text{SE}(\hat{c})}
\]
with $\hat{c}$ being the ordinary least squares estimator of $c$ and
$\text{SE}(\hat{c})$ being its estimated standard error. The corresponding quantiles can be found in \cite{Ham-1994} where in our case the standard normal distribution must be used. In case the potential cointegration relation $\beta'  X_{t}$ is known the ADF-test provides a simple method for cointegration testing. We then apply the ADF-test to the residuals $Y_t =
\beta'  X_{t-1}$. Due to $a \neq 0$ it is not necessary to include $\beta_0$.

The ADF-test can also be used to test the hypothesis \textit{integrated process} vs. \textit{trend stationarity}. An alternative is to use the KPSS-test by \cite{Kwi-1992} where the hypothesis are exchanged. A disadvantage of the ADF-test is the implicit common factor restriction which
is imposed when the ADF-test is used. The test looses power if the restriction is not
satisfied (\cite{Kre-1992}). An alternative to the ADF-test is the Wald test which tests for cointegration via the significance of the adjustment coefficients $\alpha$ (\cite{Hor-1995}).


\section{The cointegration approach to phase
synchronization}\label{sec:newdefinition}

\noindent \textbf{The cointegration model for phase processes}
\medskip

\noindent We now use the concept of cointegration for a stochastic definition of phase synchronization. The key idea is that in a stochastic context phase synchronization can be described in terms of stationarity of the phase-differences, i.e. by cointegration. That is we propose to replace the fixed deterministic bound in (\ref{weaklockingcondition}) by a stochastic bound: The difference may even get large (with small probability) but the fact that the difference is stationary will always force it back to the equilibrium.

Before we give the definition we stress that here we only want to discuss models for signals whose phase increments can be regarded as stationary - that is where the unwrapped phases are integrated processes with a deterministic linear trend (the stochastic part of the integrated process is sometimes called ``stochastic trend''). The case where the deterministic trend is zero will hardly occur in practice; the case where the stochastic trend is zero (i.e. the process is trend stationary) may occur in specific cases - see the discussion below. By integrated we always mean I(1)-integrated in terms of the cointegration literature.

\begin{defn}[Stochastic Phase Synchronization]
$d$ oscillators with phase processes  \linebreak $\phi_t =(\phi_t^{(1)},\ldots,\phi_t^{(d)})'$ are called \underline{stochastically phase synchronized} of order $r$ with $1 \le r \le d-1$, if all processes are integrated  and there exist $r$ linearly independent vectors $\beta_j$ such that the $\beta_j'\phi_t$ are stationary (i.e. $\phi_t$ is cointegrated with rank~$r$). In the case of the VEC-model (\ref{Pi-representation}) this means that ${\rm rank}(\Pi)=r$ and $\Pi = \alpha \beta'$ as in (\ref{VEC-representation}) with $\beta=(\beta_1,\ldots,\beta_r)$. The $\beta_j'\phi_t$ are up to a constant and up to some non-identifiability the phase synchronization relations.
\end{defn}

One may also use the term \textit{stochastically $\beta$-phase synchronized} of order $r$ if the processes are cointegrated of rank $r$ with cointegrating matrix/vector $\beta$.

We now briefly discuss the different orders $r$ of phase synchronization which correspond to the rank of the matrix $\Pi$ in (\ref{Pi-representation}):\\[4pt]
\noindent $\underline{r\!=\!0:}$ If ${\rm rank}(\Pi)\!=\!0$ there do not exist any cointegration (equilibrium) relations and the phases are integrated processes with a deterministic drift term (which for phase processes usually is different from $0$) and a stochastic drift term (meaning that the processes are integrated). Since the vector of phase increments follows a VAR-process as in (\ref{Pi-representation}) with $\Pi\!=\!0$ there is still some stochastic dependence of the stochastic drifts terms but this does not lead to synchronization of the phases.\\[4pt]
$\underline{0 \!<\! r \!<\! d:}$ In this case we are having $r$ phase synchronization relations, meaning that $r$ is a measure for the degree of synchronization. We give some examples:\\
-- \, if $r=1$ we may have e.g. the relation $\phi_t^{(1)}=\phi_t^{(2)}$  or for example the relation \linebreak \hspace*{0.48cm} $\sum_{j=1}^{d} \beta^{j} \phi_t^{(j)} - \beta_0 = 0$ (this form may hardly occur in practise);\\
-- \, if $r\!=\!2$ we may have two (linearly) \textit{independent} relations such as $\phi_t^{(1)}=\phi_t^{(2)}$  and \linebreak \hspace*{0.48cm} $\phi_t^{(3)}=\phi_t^{(4)}$, or two generalized relations as above;\\
-- \; if $r=d-1$ we have for example the relations $\phi_t^{(1)}=\ldots=\phi_t^{(d)}$.\\
Recall that this means that the d-dimensional process is stochastically fluctuating around these $r$ phase synchronization relations with some error correction mechanism as given in (\ref{VEC-representation 2}). The phase synchronization relations are not uniquely determined. This is reflected by the equation $\Pi = \alpha \beta' = \alpha {\xi'}^{-1} (\beta \xi)'$ with any regular $r \times r$ -matrix $\xi$ where now the columns of $(\beta \xi)$ are the phase synchronization relations. Instead of single phase synchronization relations we have a \textit{phase synchronization space} spanned by $\beta$ which is the row space of $\Pi$ \cite{Joh-1991}. Testing for the rank of $\Pi$ can be viewed as testing for certain subspaces of the phase synchronization space represented by
linear restrictions.

\medskip

The case $\underline{r=d}$ is a special and difficult one. $\phi_t^{(1)},\ldots,\phi_t^{(d)}$ then no longer are cointegrated and there does not exist any equilibrium around which the processes fluctuate. (\ref{MA-representation}) then means that the $d$-dimensional process is driven by $d\!-\!r\,$ integrated components. Thus in the case $r=d$  no stochastic trend terms remain, which are capable of driving the system; the process in (\ref{VEC-representation}) is stationary in this case. Furthermore, the term $\mu$ in (\ref{Pi-representation}) does not produce any trend in case ${\rm rank}(\Pi)=d$. One may investigate then whether all processes are trend stationary, i.e. processes with some intercept $\gamma$ and trend $\delta t$ plus some stationary part. This may be caused by an external ``pacemaker'' such as some daily cycle.

To determine $r={\rm rank}(\Pi)$ we recommend using Johansen's LR-test described above based on the VEC-representation (\ref{Pi-representation}). It is important to include the parameter $\mu$ since this generates the deterministic trend $\omega_0$ and the phase shift $\beta_0$ (as described above). If $\beta$ is known one may use other testing methods such as the ADF-test.

\cite{Kam-2010,Kam-2010b} also had the idea to use cointegration for phase synchronization and applied this to absence epilepsy data and to the analysis of neural data collected from primates. However their approach is different to the one presented here since they apply cointegration to the wrapped phases instead of the unwrapped phases. We regard this as not adequate since the wrapped phases can hardly be regarded as realizations of an integrated process. As a consequence the phase processes in their examples were often tested to be stationary and the rank test for cointegration did often lead to the maximal rank $r=d$ indicating stationarity of the phase processes instead of cointegration of integrated processes.

It is standard in physics to proceed as if the phases were observed directly and to ignore the effect of estimating the phases. The estimation of the phases is usually done by means of the Hilbert-transform on a segment (cf. Section~\ref{sec:phasesynchronization}). We will proceed in the following and in the examples in Section~\ref{sec:application} in the same way.

\bigskip

\noindent \textbf{Applying the method to phase processes}
\medskip

\noindent We now summarize the main steps of the method.
\medskip

\noindent \underline{1. Determination of the phases:}\\
Given the original observations of the oscillators, compute the phases processes $\phi_t^{(j)}$ \linebreak $(j=1,\ldots,d)$ by using the
Hilbert transform or another method (see Section~\ref{sec:phasesynchronization}). Unwrap the phases prior to the subsequent analysis. For specific models such as (\ref{CosineOscillator}),(\ref{GeneralOscillator}) other methods may be used.

\medskip
\noindent \underline{2. Testing for the rank of $\Pi$:}\\
Use the VAR-model as in (\ref{Pi-representation}) with $\mu \neq 0$ and conduct the Johansen LR-test as described above. We prefer the trace test in the ``top $\rightarrow$ bottom'' - version. For the choice of the order $p$ see the discussion in Appendix~2. It is also possible to test specific $\beta$ or hypothesis about $\beta$ (\cite{Joh-1995}, Chapter~7).

In the situation where the possible phase synchronization relations $\beta' \phi_t$ are known (in particular for $d=2$) one may use instead the ADF-test (with $a \neq 0$ and $b=0$) to test for $\beta' \phi_t$ the null hypothesis \textit{integrated process}  vs. \textit{stationary process}. If the test rejects we conclude to phase synchronization. In principle the use of our knowledge about $\beta$ results in a slightly higher power of the test.

\pagebreak

\noindent \underline{3. Estimation, interpretation and further testing of the model:}\\
If not already provided from step 2, fit the model (\ref{VEC-representation}) with $\beta$ from step 2 to the data and estimate all coefficients. Check for significance of all coefficients and delete those coefficients which are not significant. Calculate $\omega_0$ and $\beta_0$ as described above and write the whole model in the form
\begin{equation*} \label{}
(\Delta X_t - \omega_0) = \alpha \, \big(\beta'  X_{t-1} - \beta_0\big) + \sum_{i=1}^{p-1} \Gamma_i \,
\big(\Delta X_{t-i} - \omega_0 \big) + \eta_{t}.
\end{equation*}
with the estimated parameters - see (\ref{VECM-representation Example}). Remember that the parameters $\alpha$ and $\beta$ are not uniquely determined (only the cointegration space is unique). This means that different $\beta$ can be chosen to formulate the same equilibrium relations (see also  \cite{Jus-2006}, Section 8.5, ``The cointegration rank: a difficult choice''). For testing the coefficients one can use the t-ratio. In the case of cointegration this t-ratio follows a standard t-distribution.

\medskip
\noindent \underline{4. Testing for unidirectional coupling:}\\
An important special case of step 3 is the testing for unidirectional coupling. If phase synchronization is detected one can test $\alpha$ for the
direction of the dependency. For example in the VEC-model
(\ref{vecm1}), (\ref{vecm2}) with $\delta_1=\delta_2=0$ the direction of coupling can be investigated with the adjustment coefficients $\alpha_1$, $\alpha_2$. If, for example, a significance test suggests that $\alpha_1 \!<\! 0$ and $\alpha_2 \!=\! 0$  one can conclude that $\phi_t^{(2)}$ influences/corrects $\phi_t^{(1)}$ but not the other way round. Thus, $\phi_{t}^{(2)}$ is the
driving force of phase synchronization and $\phi_{t}^{(1)}$ is forced to adapt.

\medskip
Software for performing these steps is discussed in the Appendix~\ref{sec:Appendix2:computational}.
\begin{figure}
\centering
\includegraphics[width=400pt,
keepaspectratio]{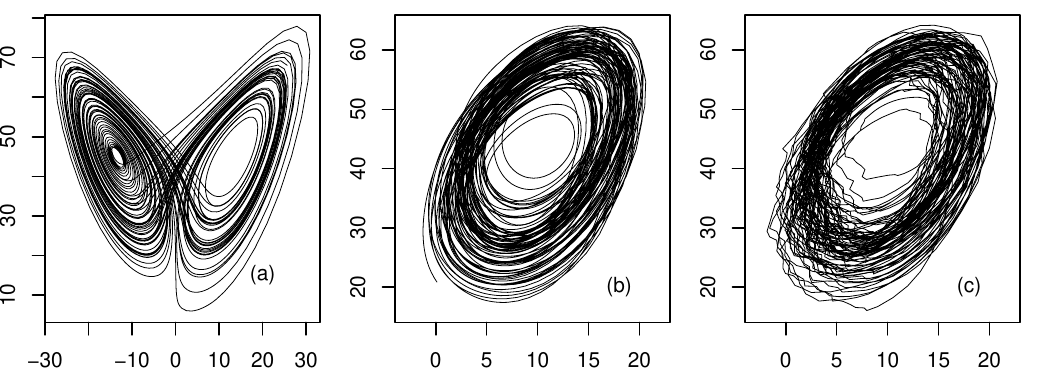}
\caption{\footnotesize
R\"{o}ssler-Lorenz system: Trajectory of the Lorenz attractor in the $xz$-plane ($z_2\,${\scriptsize vs.}$\,x_2$)
for coupling strengths $\epsilon=0$ (a) and $\epsilon=12$ (b). (c) shows
the trajectory in (b) corrupted by noise.}
\label{fig:LorenzXZProjection}
\end{figure}

\section{Examples} \label{sec:examples}

\subsection{A coupled R\"{o}ssler-Lorenz system} \label{sec:application}

As an example we analyze with the above methods an unidirectionally coupled R\"{o}ssler-Lorenz system. We proceed through steps 1-4 above. Computational details and additional modeling aspects can be found in Appendix~2.

\medskip
\noindent \underline{0. Simulation of the data:}\\
\noindent The system is defined through an autonomous R\"{o}ssler attractor with configuration
\begin{eqnarray*}
\dot{x_1} &=& -12 (y_1 + z_1) + w_1,\qquad \quad \; \\
\dot{y_1} &=& 12 \, (x_1 + 0.2 y_1),\\
\dot{z_1} &=& 12\,\big[0.2 + z_1(x_1-5.7)\big],
\end{eqnarray*}
and a driven Lorenz attractor given by
\begin{eqnarray*}
\qquad \dot{x_2} &=& 16 \,(y_2-x_2) - \epsilon \,(x_2-x_1)  + w_2,\\
\dot{y_2} &=& 45.92x_2 - y_2 - x_2 z_2,\\
\dot{z_2} &=& x_2 y_2 - 4 z_2.
\end{eqnarray*}
with system noise variables $w_1$, $w_2$ being i.i.d. $\mathcal{N}(0,0.15^2)$ distributed similar to the system analyzed in \cite{Schel-2007}. The parameters are taken from \cite{Gua-2005}. The coupling is induced through the inclusion of the $x_1$ term in the equation of $\dot{x_2}$ with coupling strength $\epsilon$. Related systems have been discussed before in \cite{Gua-2005,Qui-2000,Pal-2001,Pal-2007}. In our study coupling strengths $\epsilon$ between $8.0$ and $12.0$ are used (see below).

\begin{figure}
\centering
\includegraphics[width=400pt,
keepaspectratio]{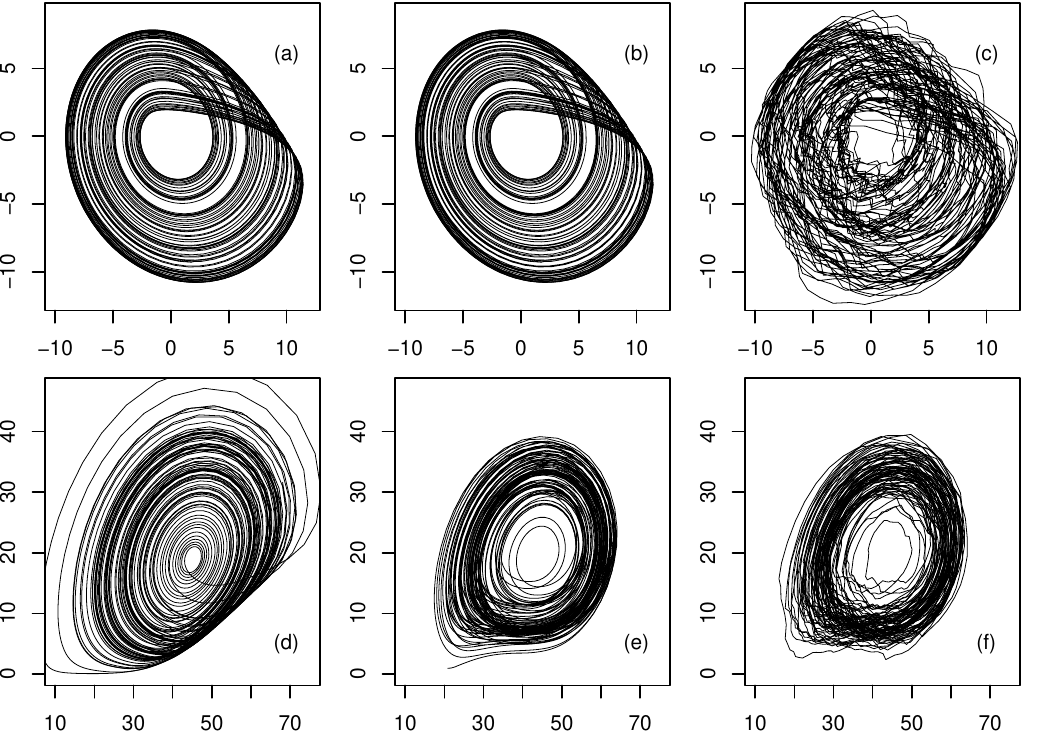}
\caption{\footnotesize
R\"{o}ssler-Lorenz system: Trajectories of the R\"{o}ssler attractor in
the $xy$-plane (a)-(c) ($y_1${\scriptsize vs.}$x_1$) and Lorenz attractor in the $uz$-plane (d)-(f) ($z_2${\scriptsize vs.}$u_2$) for
coupling strengths $\epsilon=0$ (a),(d) and $\epsilon=12$ (b),(e) (as a
result of the unidirectional coupling (a) and (b) are identical). (c)/(f) show
the trajectories in (b)/(e) corrupted by noise.}
\label{fig:RoesslerLorenzProjections}
\end{figure}

The solution is approximated by integration from 0 to 50 with step size 0.01 by using a fourth-order Runge-Kutta algorithm (cf. \cite{Pre-1992}) leading to 5,000 data points. Note, that the above notation is common in physics but in a mathematical sense not correct (there are derivatives on the left-hand side and discrete time noise on the right-hand side). The precise meaning is, that in the discrete approximation with the Runge-Kutta algorithm the right hand side is used at each gridpoint in time with additional discrete noise. To make the setting more realistic stationary observation noise is added to the oscillators: To the computed values $x_{i,t},y_{i,t},z_{i,t}$ at time $t$  we add $u_t^i = 0.9 u_{t-1}^i + v_t^i$
with $v_t^i \sim \mathcal{N}(0,0.1)$.

The trajectory of the Lorenz attractor (in the $xz$-plane) for $\epsilon=0$
(no coupling) and $\epsilon=12$ (coupling) are given in Figure~\ref{fig:LorenzXZProjection}.
It can be observed that, in contrast to the R\"{o}ssler attractor (see
Figure~\ref{fig:RoesslerLorenzProjections} (a)-(c)),  the Lorenz attractor has
two rotation centers in the uncoupled case. Therefore, the phase of the Lorenz attractor is usually defined in the $uz$-plane where $u= \sqrt{x^2 + y^2}\,$ (cf. \cite{Pik-1997}). In the $uz$-plane only one rotation center exists which is illustrated in Figure~\ref{fig:RoesslerLorenzProjections} (d)-(f).

\medskip
\noindent \underline{1. Determination of the phases:}\\
\noindent We apply the Hilbert transform to estimate the phases as
described in Section~\ref{sec:phasesynchronization}. The Hilbert transform is applied to the $x$ coordinate of the R\"{o}ssler attractor and to the $u$ coordinate of the Lorenz attractor, using a rolling window of 1,000 data points. After phase estimation we have removed the first and last
500 data points finally leading to 4,000 data points.

\begin{figure}
\centering
\includegraphics[width=430pt,keepaspectratio]{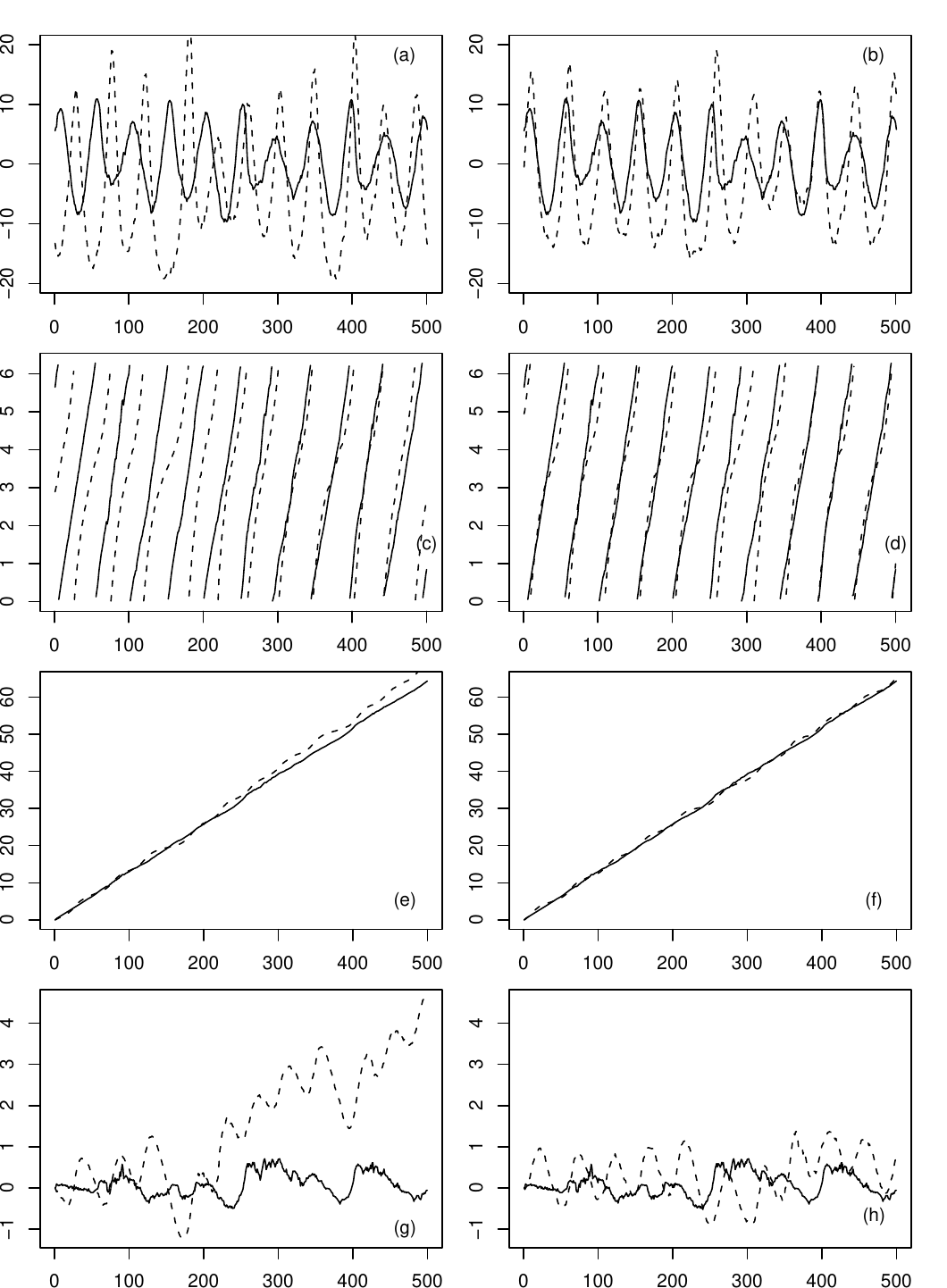}
\caption{\footnotesize
R\"{o}ssler-Lorenz system: 500 data points of the $x$ coordinate of the
R\"{o}ssler attractor (solid lines) and of the $u$ coordinate of Lorenz
attractor (dashed lines) for $\epsilon=0$ (a) and $\epsilon=12$ (b).
The data points are corrupted by noise. (c), (d) give the
corresponding wrapped phase estimates, (e), (f) the unwrapped phases, and (g), (h) the stochastic trend of the unwrapped phases.}
\label{fig:RoesslerLorenzPhases}
\end{figure}

Figure~\ref{fig:RoesslerLorenzPhases} (a),(b) show 500 data points of the $x$
and $u$ coordinate of the R\"{o}ssler and Lorenz attractors, respectively, for the
uncoupled case ($\epsilon=0$) and the coupled case ($\epsilon=12$). In (c),(d) the
wrapped phase estimates are given. It can be seen that in the uncoupled case the mean frequency of the R\"{o}ssler system is smaller than the mean frequency of the Lorenz system. In fact, for
the given configuration the natural frequencies of the R\"{o}ssler system and
the Lorenz system are $\omega_{R} = 0.129$ and $\omega_{L} = 0.137$,
respectively (see below), i.e. the natural frequencies differ significantly in the
uncoupled system. In (e),(f) the unwrapped phases are plotted and in (g),(h) the ``stochastic trend'' (in terms of cointegration), i.e. the unwrapped phases minus $\,0.129 t\,$ are plotted (the latter being the deterministic trend of the R\"{o}ssler-attractor and the coupled system as determined below). The shape of the stochastic trend in (g) already indicates no cointegration (phase synchronization) while in (h) it indicates cointegration. The oscillation of the phase of the Lorenz attractor is due to the fact that the Lorenz attractor has a slightly steeper ascent than a descent (this can be overcome by choosing a higher AR-model order - see the comments in Appendix~2).

\medskip
\noindent \underline{2. and 2a. Testing for phase synchronization:}\\
We first apply the Johansen rank test in the ``top $\rightarrow$ bottom'' - version where we set the AR-order for simplicity to $p=2$. For example for $\epsilon=12$ the hypothesis $\mathcal{H}(0)$ is clearly rejected with a LR-value of $92.44$ ($5\%$-critical value: $15.41$) and the hypothesis $\mathcal{H}(1)$ is clearly not rejected (i.e. accepted) with a LR-value of $0.01$ ($5\%$-critical value: $3.84$). This value is remarkably small, indicating the good fit of the model. Thus the test reveals ${\rm rank} \, \Pi = 1$, and we conclude to phase synchronization. Furthermore, the procedure also detects the $1\!:\!1$ - relation of the phase synchronization from the corresponding eigenvector.

\begin{figure}
\centering
\includegraphics[width=400pt,keepaspectratio]{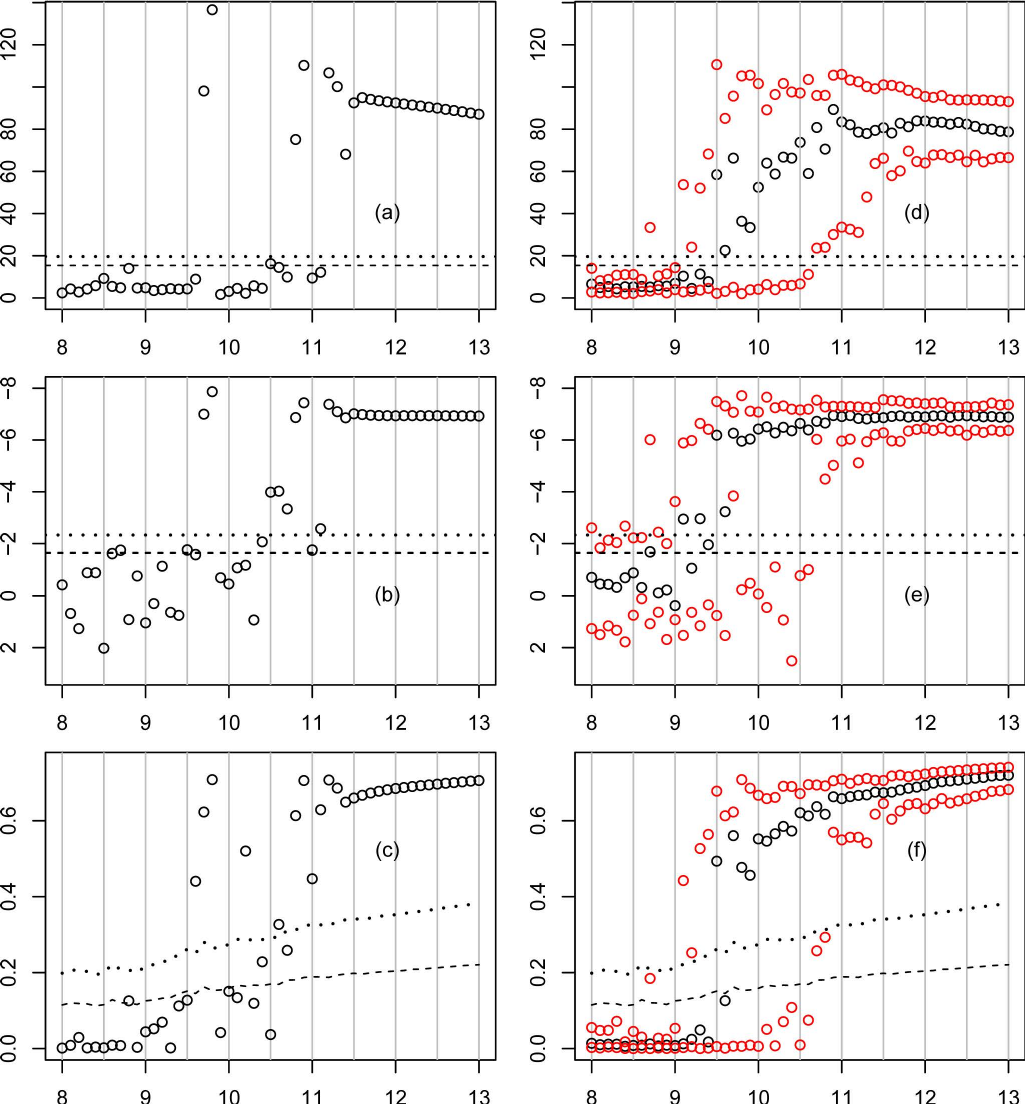}
\caption{\footnotesize
R\"{o}ssler-Lorenz system: (a),(d) values of the LR-test for $\mathcal{H}(0)$; (b),(e) values of the ADF-test; (c),(f) values of the phase synchronization index $\hat{R}^{2}$. (a)-(c) show the results of one simulation for each coupling strength $\epsilon$ between 8 and 13; (d)-(f) show the
median and the 10\% and 90\% quantiles of a larger simulation study. The dashed lines and the dotted lines show the significance levels
$\alpha=0.05$ and $\alpha=0.01$, respectively.}
\label{fig:ComparisonToPSIndex}
\end{figure}

Figure~\ref{fig:ComparisonToPSIndex}~(a) shows the values of the $\mathcal{H}(0)$ - test statistic for different $\epsilon$ (but the same values for the noise variates). The dashed line and the
dotted line are the critical values $15.41$ and $19.62$ for the significance levels
$0.05$ and $0.01$, respectively (c.f. \cite{Jus-2006}, Appendix A, Case 3). For those $\epsilon$ where the test rejects the hypothesis $\mathcal{H}(1)$ was tested afterwards and in all cases not rejected. Thus the plotted value of the $\mathcal{H}(0)$ - statistic
determined the decision \textit{phase synchronization} vs. \textit{no phase synchronization}. The values for $\epsilon > 11.1$ are all highly significant corresponding to \textit{phase synchronization} while the values for $\epsilon \le 11.1$ are nearly all not significant indicating \textit{no phase synchronization}.

\begin{figure}
\centering
\includegraphics[width=450pt,keepaspectratio]{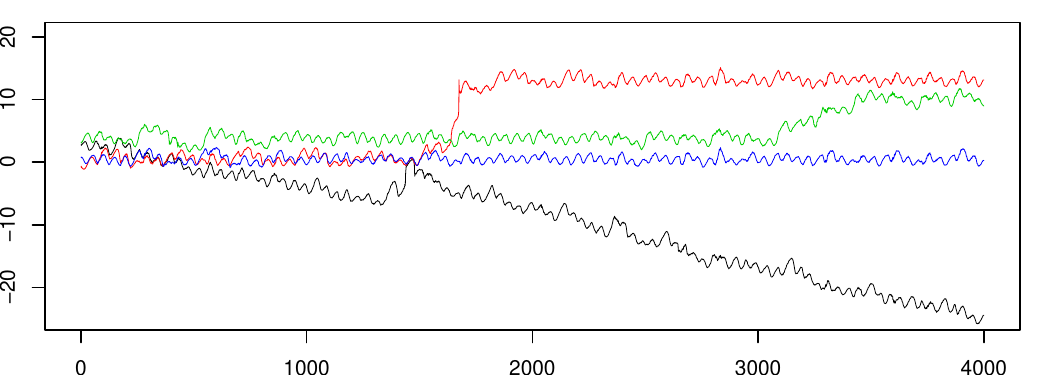}
\caption{\footnotesize
R\"{o}ssler-Lorenz system: Phase differences $\phi_t^{(1)}-\phi_t^{(2)}$ for coupling strength
$\epsilon=0$ (black), $\epsilon=9.6$ (green), $\epsilon=10.2$ (red) \, and $\epsilon=12$ (blue).}
\label{fig:RoesslerLorenzPhaseDiff}
\end{figure}

In Figure~\ref{fig:ComparisonToPSIndex}~(c) the phase synchronization index $\hat{R}^{2}$ is plotted. The associated critical values are computed as explained in \cite{Schel-2007}. The results
of the two tests agree apart from $\epsilon = 9.6, 10.2, 11.0, 11.1$ (and $10.6$ as a ``borderline''-case).
The difference can in all 4 cases be explained by $2\pi$-jumps of the phase - differences which are penalized by the LR-test but not penalized by the $\hat{R}^{2}$-statistic. The situation is highlighted in Figure~\ref{fig:RoesslerLorenzPhaseDiff} where we have plotted the estimated phase differences $\phi_t^{(1)}\!\!-\!\phi_t^{(2)}$ for $\epsilon = 0$, $9.6$, $10.2$ and $12$ (the situation for $11.0, 11.1$ being similar to the cases $9.6$ and $10.2$). We think it is a matter of taste whether one calls the cases $9.6, 10.2$ phase synchronized (see the comments below).

We have also used the testing procedure as described under 2a. based on the ADF-test. Here we have to specify the possible phase synchronization relation in advance. Based on the plot in Figure~\ref{fig:RoesslerLorenzPhases}~(b) we have decided to test for a $1\!:\!1$ synchronization relation, that is we have applied the ADF-test to $\phi_t^{(1)}\!\!-\!\phi_t^{(2)}$. The results are given in Figure~\ref{fig:ComparisonToPSIndex}~(b) with the vertical axis being upside down. The dashed line and the dotted line are the critical values $-1.65$ and $-2.33$ associated with the significance levels $0.05$ and $0.01$, respectively (from the standard normal distribution - see \cite{Ham-1994}, p. 529, Case 3). The results coincide with the LR-test (with a slight difference for $\epsilon = 10.5, 10.6, 10.7$).

Figure~\ref{fig:RoesslerLorenzPhases}~(d)-(f) show the median and the 10\% and 90\% quantiles of a larger number of simulations where the cases with phase jumps have been removed. The plots and the severe limitations of this simulation are discussed in detail in the last paragraph of this section. As a positive outcome the plots confirm that the tests behave similarly. They also indicate that the ADF-test has a higher power than the Johansen rank test for values of $\epsilon$ between $9.5$ and $10.5$ which is not surprising since the ADF-test uses as an additional information the form of the cointegration relation (however, the true power of the tests is unknown and cannot be estimated from this simulation).

It is remarkable that the LR-test and the ADF-test as two standard tests from the theory of cointegration perform similar to the $\hat{R}^{2}$-test which has been explicitly tailored for phase synchronization. The tests are only different for processes with phase jumps which we regard more as a feature of chaotic oscillators than of real data - see also the discussion in Section~\ref{sec:conclusion}. We also mention that we do not regard the phase synchronization index $\hat{R}^{2}$ from Figure~\ref{fig:ComparisonToPSIndex}~(c) as some `gold-standard'. Instead we think, that for example a (properly adjusted) LR-test in an adequate model would be a  better choice.


\medskip

\noindent \underline{3. and 3a. Estimation, further testing, and unidirectional coupling:}\\
Given that $\phi_t^{(1)}$, $\phi_t^{(2)}$ are cointegrated with cointegrating vector
$\beta = (1,-1)'$ we estimate the VEC-model (\ref{vecm1}),(\ref{vecm2}) to analyze
the joint dynamics with the aim to uncover the directional coupling.
Here we restrict to the case $\epsilon = 12$ from Figure~\ref{fig:RoesslerLorenzPhases} (a)-(c).
The estimation results are given in Table~\ref{table:estvecm}.

\bigskip

\begin{table}[th]
\centering
\begin{tabular}{c|cccc}
\hline
Parameter & Estimate & Std. Error & t-ratio & p-value\\
\hline
\hline
$\alpha_1$ & $0.0009$ & $0.0015$ & $0.62$ & $0.53$\\
$\gamma_1$ & $0.059$ & $0.016$ & $3.71$ & $0.0002$\\
$\delta_1$ & $0.024$ & $0.014$ & $1.70$ & $0.089$\\
$\mu_1$ & $0.1177$ & $0.0029$ & $41.12$ & $<0.0001$\\
\hline
$\alpha_2$ & $0.0110$ & $0.0011$ & $9.65$ & $<0.0001$\\
$\gamma_2$ & $0.726$ & $0.011$ & $68.38$ & $<0.0001$\\
$\delta_2$ & $0.028$ & $0.012$ & $2.38$ & $0.017$\\
$\mu_2$ & $0.0262$ & $0.0022$ & $12.11$ & $<0.0001$\\
\hline
\end{tabular}
\caption{\footnotesize Estimated VEC-model for the R\"{o}ssler-Lorenz system.}
\label{table:estvecm}
\end{table}
In the equation of $\Delta \phi_t^{(1)}$ only
$\mu_1$ and $\gamma_1$ have a significant t-ratio. In the equation of $\Delta \phi_t^{(2)}$ all parameters except $\delta_2$ are highly significant. Since $\delta_1$ and  $\delta_2$ as well as the cross-correlations of the residuals with $\text{Corr} \, (\eta^{(1)}_t,\eta^{(2)}_t) \!= 0.022$  are not significantly different from zero, the dependency of the phases is explained in the model purely by the error correction mechanism. Since $\alpha_1=0$ and $\alpha_2 \!>\! 0$ the VEC-model reliably identifies the direction of coupling of the original R\"{o}ssler-Lorenz system.

Finally, a reestimation of the model with only the significant parameters
yields
\begin{eqnarray}
\label{fittedvecm1}
\Delta \phi_t^{(1)} &=& 0.060 \, \Delta \phi_{t-1}^{(1)} + 0.121 + \eta^{(1)}_t,\\
\label{fittedvecm2}
\qquad \Delta \phi_t^{(2)} &=& 0.011 \, \big(\phi_{t-1}^{(1)} - \phi_{t-1}^{(2)}\big)
 + 0.727 \, \Delta \phi_{t-1}^{(2)} + 0.030 + \eta^{(2)}_t.
\end{eqnarray}
Note also the more informative representation (\ref{VECM-representation Example}) below and its discussion.

We now discuss the estimated system in terms of the modified VEC-representation (\ref{VEC-representation 2}) and the MA-representation (\ref{MA-representation}). At the same time this system is a nice illustration of the theoretical setting presented in Section~\ref{sec:cointegrationGranger} (and a nice example for a cointegrated system). With the above estimates we obtain
\begin{equation*} \label{}
\beta =  \left(
\begin{matrix}
1 \\
-1
\end{matrix}
\right), \quad \beta_{\bot} = \left(
\begin{matrix}
1 \\
1
\end{matrix}
\right), \quad \alpha = \left(
\begin{matrix}
0 \\
0.011
\end{matrix}
\right), \quad \alpha_{\bot} = \left(
\begin{matrix}
1 \\
0
\end{matrix}
\right), \quad \mu = \left(
\begin{matrix}
0.121 \\
0.030
\end{matrix}
\right).
\end{equation*}
This leads with straightforward calculation to the other values
\begin{equation*} \label{}
\Gamma_1 =  \left(
\begin{matrix}
0.06 & \!0\\
0 & \!0.73
\end{matrix}
\right)\!, \;\, \Gamma =  \left(
\begin{matrix}
0.94 & \!0\\
0 & \!0.27
\end{matrix}
\right)\!, \;\,  C = 0.94^{-1} \!\left(
\begin{matrix}
1\\
1
\end{matrix}
\right) \! \big(1,0\big), \;\,  \omega_0 = C \mu =0.129 \left(
\begin{matrix}
1\\
1
\end{matrix}
\right),
\end{equation*}

\noindent and $\beta_0 = 0.49$. With these values we obtain the VEC-representation %
\begin{equation} \label{VECM-representation Example}
\Bigg[ \!\bigg(
\begin{matrix}
\Delta \phi_{t}^{(1)}\\
\Delta \phi_{t}^{(2)}
\end{matrix}
\bigg) \!-\! \bigg(
\begin{matrix}
0.129\\
0.129
\end{matrix}
\bigg) \!\Bigg] \!= \!\Big(
\begin{matrix}
0 \\
0.011
\end{matrix}
\Big)  \Big(\phi_{t-1}^{(1)} - \phi_{t-1}^{(2)} - 0.49 \Big) +  \left(
\begin{matrix}
0.06 & 0\\
0 & 0.73
\end{matrix}
\right) \!
\Bigg[\!\bigg(
\begin{matrix}
\Delta \phi_{t-1}^{(1)}\\
\Delta \phi_{t-1}^{(2)}
\end{matrix}
\bigg) \!-\! \bigg(
\begin{matrix}
0.129\\
0.129
\end{matrix}
\bigg) \!\Bigg] + \eta_{t} \quad
\end{equation}
and the MA-representation
\begin{equation} \label{MA-representation Example}
\phi_t = \left(
\begin{matrix}
1\\
1
\end{matrix}
\right) \sum_{i=1}^{t} \eta_{i}^{(1)} + \left(
\begin{matrix}
0.129\\
0.129
\end{matrix}
\right) t + C^{*}(L) (\eta_t
 + \mu) + \phi_0.
 \end{equation}
(\ref{VECM-representation Example}) has the meaning that the phase process is pulled towards the equilibrium defined by $\beta'  \phi_{t} \!-\! \beta_0 = \phi_{t}^{(1)} \!-\! \phi_{t}^{(2)} \!- 0.49 = 0$ with the force $\alpha = \binom{0}{0.011}$ which gets active as soon as the process leaves the equilibrium. In this case $\alpha_1=0$ i.e. the correction is on process $2$ only. This  means that the unidirectional coupling between the R\"{o}ssler and the Lorenz system has been detected correctly.

The equilibrium $\,\phi_{t}^{(1)} \!-\! \phi_{t}^{(2)} \!- 0.49 = 0\,$ means that the phase of the R\"{o}ssler system is ahead of the phase of the Lorenz system in the average by $0.49$. This is  confirmed by the plot in Figure~\ref{fig:RoesslerLorenzPhases} (d).

The representation (\ref{MA-representation Example}) has the meaning that both phases are mainly pushed in direction $\binom{1}{1}$ - both with a deterministic and a random walk component. The stationary component $ C^{*}(L) (\eta_t + \mu)\,$ (whose precise form is complicated) acts as a disturbance component pushing the system with little shocks again and again out of the equilibrium.

For comparison, we also fit the phase model $\Delta \phi_t = \gamma \Delta \phi_{t-1} + \mu + \eta_t$ to each oscillator in the uncoupled case $\epsilon=0$ which gives
\begin{eqnarray}
\label{fittedphasemodel1}
\big(\Delta \phi_t^{(1)} - 0.129\big) &=& 0.060 \, \big(\Delta \phi_{t-1}^{(1)} - 0.129\big) + \eta^{(1)}_t,\\
\label{fittedphasemodel2}
\big(\Delta \phi_t^{(2)} - 0.137\big) &=& 0.735 \, \big(\Delta \phi_{t-1}^{(2)} - 0.137) + \eta^{(2)}_t,\qquad
\end{eqnarray}
((\ref{fittedvecm1}) and (\ref{fittedphasemodel1}) are of course identical). As mentioned above the natural frequencies of the R\"{o}ssler and the Lorenz system therefore are $\omega_{R} = 0.129$ and $\omega_{L} = 0.137$, respectively, i.e. the natural frequencies differ significantly in the
uncoupled system, while in the coupled system they both are $\omega = 0.129$ (remember that
$\omega t$ is the deterministic part of the trend).

One effect should be kept in mind: The prior application of the Hilbert transform leads to phases which are smoother than the original ``true'' ones. This effect needs to be investigated in the future (see Appendix 1). As a consequence the estimated variances of $\eta_t$ and $\phi_t$ will usually be smaller than the true ones. In (\ref{fittedphasemodel1}), (\ref{fittedphasemodel2}) one finds $\text{Var} \, \eta^{(1)}_t \!= 0.0034$ and $\text{Var} \,  \eta^{(2)}_t \!= 0.0025$ leading to $\text{Var} \, \big(\Delta \phi_t^{(1)}\big) \!\!= 0.0034$ and $\text{Var} \, \big(\Delta\phi_t^{(2)}\big) \!\!= 0.0054$. This means that the
Lorenz attractor has a more varying frequency than the R\"{o}ssler attractor and it is likely that this also holds for the true frequencies.

Perhaps the most important aspect of this fitted VEC-model is that the direction of coupling is detected correctly. In Figure~\ref{fig:RoesslerLorenzPhases}~(d)  in total $1203$ series have been tested positively as being phase-synchronized with the LR-test. Out of these, $93.6$\% have been tested by the VEC-model as having unidirectional coupling. This decision was characterized by a p-value larger than $0.1$ for $\alpha_1$ and a p-value smaller than $0.01$ for $\alpha_2$. This means that in more than $93.6$\% cases of the synchronized one sided-coupled R\"{o}ssler-Lorenz system a testing value smaller than the $90$\%-quantile of the stochastic cointegration-model was obtained for $\alpha_1$. This also confirms that the stochastic model is reasonable.

We finally mention the limitations of sampling from the R\"{o}ssler-Lorenz system, and of the plots in Figure~\ref{fig:RoesslerLorenzPhases}~(d)-(f) for the purpose of this paper: As a chaotic oscillator the R\"{o}ssler-Lorenz system is very sensitive to the starting values and also to the random inputs $w_1$ and $w_2$ leading under the simulation roughly to 4 different path-types: synchronized oscillators, non-synchronized oscillators without phase jumps (over the whole segment), oscillators with phase jumps, segmentwise combinations of the former three cases. Even for the fist two cases the approximation with the cointegration model of this paper leads to very different parameter values, and therefore to different models from a stochastic viewpoint. For this reason simulations from the R\"{o}ssler-Lorenz system are not suitable for e.g. judging the properties of estimates or tests in detail (with the above remarks on the higher power of the ADF-test and the suitability for detecting unidirectional coupling as an exception). In detail for Figure~\ref{fig:RoesslerLorenzPhases}~(d)-(f) we have done 100 simulations for each $\epsilon$ (with the same noise variates). Out of a total of 5100 paths those paths with phase jumps and segmentwise changing structure have been removed by ad-hoc testing and eye-inspection, leading to only $1530$ cases (for example for $\epsilon=8.8,9.2,9.4,9.6,9.7$ less than $14$ values remained) which is the reason for the erratic behavior of the quantiles in Figure~\ref{fig:RoesslerLorenzPhases}~(d)-(f). Of course the subjective method of excluding the other cases is questionable and it makes no sense to repeat the procedure for a larger number of simulations. To summarize the cointegration model of this paper is not a statistical model for the R\"{o}ssler-Lorenz system as a whole but only for single paths of it, and moreover for the detection of phase synchronization of other time series not stemming from chaotic oscillators.

\subsection{Synchronization of electrochemical oscillators} \label{sec:application2}
We now investigate with the above methods phase synchronization in a chaotic process in electrochemistry. The system is the electrodissolution of nickel in sulfuric acid. The electrodissolution takes place on two nickel wires, and the rate of the dissolution
(corrosion rate), measured as the current, exhibits chaotic behavior - see \cite{Kis-2002}. The
oscillators are coupled through a set of serial and parallel resistors. Such  coupling induces a bidirectional, electrical
interaction between the electrochemical processes.  For more details and a schematic diagram of the experimental apparatus see \cite{Kis-2002}.

\begin{figure}
\centering
\includegraphics[width=450pt,keepaspectratio]{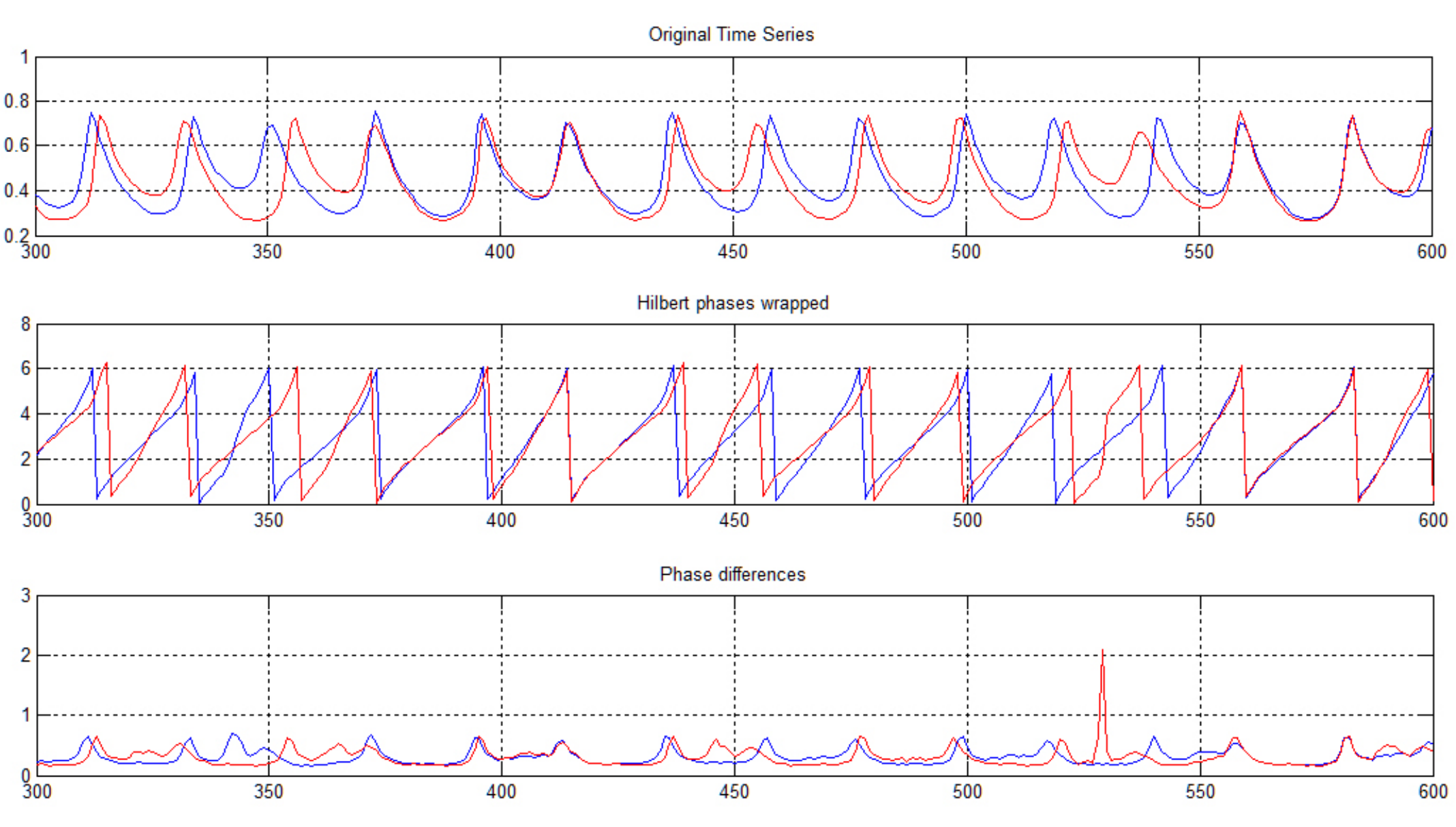}
\caption{\footnotesize
Currents at 2 electrodes of an electrochemical oscillator: The figures show the original currents, the Hilbert estimate of the phases before unwrapping, and the phase differences $\Delta \phi_t$ for each phase process (the first component is in blue, the second in red).}
\label{fig:ElectrochemicalOscillator1}
\end{figure}

The phase synchronization of the chaotic chemical process was established using non-statistical methods (an entropy based measure of
the cyclic phase difference distribution) in
\cite{Kis-2002} and \cite{Kis-2005}. In this section, we explore the characteristics of the synchronization process using the cointegration technique. The Hilbert transform is applied to estimate the phases of the two-dimensional time series, and afterwards more than 5 seconds of data points are removed at the beginning and the end. A data set is considered with data acquisition rate of 25 Hz  leading to about 21 data points per cycle. The whole series then consists of 3750 2-dimensional data points. In order to test the methods on a smaller data set we split this in 6 segments of equal length each consisting of 625 data points. The results on each of the 6 segments are essentially the same and for clarity we therefore report only the results for one segment (instead of displaying summary statistics etc.). Furthermore, we mention some results for the whole stretch of 3750 data points. Figure~\ref{fig:ElectrochemicalOscillator1} shows 300 data points of the two oscillators from the investigated segment (the first component is in blue, the second in red), the estimated phases before unwrapping, and the phase differences $\Delta \phi_t$. The plot reveals that the phases and the phase-differences have an in-cycle fluctuation which comes from the phase slowing down at the minimum of the oscillations and speeding up at the maximum.

The AR-order $p$ in (\ref{VEC-representation}) is chosen by inspection of the significant lags in the model fit below leading to the choice $p=6$. This order is necessary because of the above mentioned in-cycle fluctuations. The red peak for the phase differences is due to the fact that the corresponding cycle is rather short resulting in large phase differences.
As above the Johansen LR test and the ADF test are applied to test for cointegration of the phases and therefore for phase synchronization of the original series. The hypothesis $\mathcal{H}(0)$ is clearly rejected with a LR-value of $52.68$ ($5\%$-critical value: $17.95$) and the hypothesis $\mathcal{H}(1)$ is clearly not rejected (i.e. accepted) with a LR-value of $0.02$ ($5\%$-critical value: $8.18$). As above this value is very small, indicating the good fit of the model.
Thus the test reveals ${\rm rank} \, \Pi = 1$, and we conclude to phase synchronization. Furthermore, the procedure also detects the $1\!:\!1$ - relation of the phase synchronization from the corresponding eigenvector. For the long series consisting of 3750 data the LR-value for $\mathcal{H}(0)$ is even $321.94$ and for $\mathcal{H}(1)$ it is $0.03$.
The ADF-test leads with a testing value of $-6.398$ and a $1\%$-critical value
of $-3.43$ to the rejection of the hypothesis that the phase are not cointegrated. The same result is obtained with a testing value of $-17.9464$ for the long series, i.e. the ADF-test also confirms phase synchronization.
Given that $\phi_t^{(1)}$, $\phi_t^{(2)}$ are cointegrated with cointegrating vector
$\beta = (1,-1)'$ we estimate the VEC-model (\ref{VEC-representation}) with order $p=6$ to analyze
in more detail the joint dynamics. Different to the results in Section~\ref{sec:application} several AR-parameters are significant. The estimated coefficients are displayed in Table~\ref{table:estvecm2}.
\bigskip
\begin{table}[th]
\centering
\begin{tabular}{c|cccc}
\hline
Parameter & Estimate & Std. Error & t-ratio & p-value\\
\hline
\hline
$\alpha_1$ & $\!\!-0.011$ & $0.003$ & $\!\!-3.72$ & $\,0.0002$\\
$\mu_1$   & $\;\,0.073$ & $0.008$ &  $\;\,8.96$ & $<0.0001$\\
$\Gamma_{\scriptsize{\mbox{1}}_{11}}$ & $\;\,1.565$ & $0.040$ & $38.65$ & $<0.0001$\\
$\Gamma_{\scriptsize{\mbox{2}}_{11}}$ & $\!\!-1.150$ & $0.075$ & $\!\!\!\!\!-15.36$ & $<0.0001$\\
$\Gamma_{\scriptsize{\mbox{3}}_{11}}$ & $\;\,0.567$ & $0.086$ & $\;\,6.61$ & $<0.0001$\\
$\Gamma_{\scriptsize{\mbox{4}}_{11}}$ & $\!\!-0.206$ & $0.076$ & $\!\!-2.71$ & $0.007$\\
\hline
$\alpha_2$ & $\;\,0.033$ & $0.006$ & $\;\,5.62$ & $<0.0001$\\
$\mu_2$ & $\;\,0.102$ & $0.017$ & $\;\,5.97$ & $<0.0001$\\
$\Gamma_{\scriptsize{\mbox{1}}_{21}}$ & $\;\,0.427$ & $0.084$ & $\;\,5.06$ & $<0.0001$\\
$\Gamma_{\scriptsize{\mbox{1}}_{22}}$ & $\;\,0.740$ & $0.040$ & $18.57$ & $<0.0001$\\
$\Gamma_{\scriptsize{\mbox{2}}_{21}}$ & $\!\!-0.444$ & $0.156$ & $\!\!-2.84$ & $0.005$\\
$\Gamma_{\scriptsize{\mbox{2}}_{22}}$ & $\!\!-0.142$ & $0.050$ & $\!\!-2.85$ & $0.005$\\
$\Gamma_{\scriptsize{\mbox{5}}_{21}}$ & $\!\!-0.207$ & $0.088$ & $\!\!-2.35$ & $0.019$\\
\hline
\end{tabular}
\caption{\footnotesize Estimated VEC-model for the electrochemical oscillators. Note that all other coefficients $\Gamma_{\scriptsize{\mbox{1}}_{ij}},\ldots,\Gamma_{\scriptsize{\mbox{5}}_{ij}}$ are not significantly different from $0$.}
\label{table:estvecm2}
\end{table}

As expected there is no unidirectional coupling which is reflected by $\alpha_1 \!<\! 0$ and $\alpha_2 \!>\! 0$. It is however remarkable that $\alpha_2$ is 3-times as large as $\alpha_1$. On the whole segment with 3750 data we obtain as estimates $\alpha_1=-0.012$ and $\alpha_2=0.026$, i.e. in total $\alpha_2$ is more than twice as large as $\alpha_1$. The same holds on all 6 segments and also when trying different AR-orders. In addition the AR-coefficients $\Gamma_{\scriptsize{\mbox{k}}}$ indicate an influence from series 2 to series 1. Summarizing, the force towards the phase equilibrium is much stronger from series 2 on series 1 than vice versa. These results
thus show that while the physical form of the coupling is symmetrical (through differences in the electrode potentials~\cite{Kis-2002}), the effective
coupling on the phase dynamics could develop asymmetries, likely due to the heterogeneities in the local dynamics.

We finally discuss the estimated system in terms of the modified VEC-representation (\ref{VEC-representation 2}). The above table yields
\begin{equation*} \label{}
\beta =  \left(
\begin{matrix}
1 \\
-1
\end{matrix}
\right), \quad \beta_{\bot} = \left(
\begin{matrix}
1 \\
1
\end{matrix}
\right), \quad \alpha = \left(
\begin{matrix}
-0.011\\
\;\;\,0.033
\end{matrix}
\right), \quad \alpha_{\bot} = \left(
\begin{matrix}
0.033 \\
0.011
\end{matrix}
\right), \quad \mu = \left(
\begin{matrix}
0.073\\
0.102
\end{matrix}
\right)
\end{equation*}
plus the values for the $\Gamma_k$ (not listed here) leading with $\;\Gamma \!=\! I_{p}-\sum_{i=1}^{p-1} \Gamma_{i}$,  $\;C \!=\! \beta_{\bot} \Big[\alpha_{\bot}' \Gamma \beta_{\bot}\Big]^{\!-1} \!\alpha_{\bot}'$, $\bar{\alpha} = \alpha (\alpha' \alpha)^{-1}$ to
\begin{equation*} \label{}
\omega_0 = C \mu = 0.302 \left(
\begin{matrix}
1\\
1
\end{matrix}
\right) \qquad \mbox{and} \qquad \beta_0 = \bar{\alpha}' \big[ \Gamma C - I_d \big] \mu = 0.077
\end{equation*}
meaning that the 'true' estimated equilibrium is $\phi_t^{(1)}-\phi_t^{(2)}-0.077=0$. This leads to the VEC-representation
\begin{equation} \label{VECM-representation Example}
\Bigg[ \!\bigg(
\begin{matrix}
\Delta \phi_{t}^{(1)}\\
\Delta \phi_{t}^{(2)}
\end{matrix}
\bigg) \!-\! \bigg(
\begin{matrix}
0.302\\
0.302
\end{matrix}
\bigg) \!\Bigg] \!= \!\Big(
\begin{matrix}
-0.011 \\
\;\;\,0.033
\end{matrix}
\Big)  \Big(\phi_{t-1}^{(1)} - \phi_{t-1}^{(2)} - 0.077 \Big) +  \; \mbox{higher order AR-terms} \; + \eta_{t}
\end{equation}
which demonstrates the error correcting effect on the first and second component if the oscillators are out of equilibrium. For completeness we mention that the corresponding fit to the long series leads to
\begin{equation} \label{VECM-representation Example}
\Bigg[ \!\bigg(
\begin{matrix}
\Delta \phi_{t}^{(1)}\\
\Delta \phi_{t}^{(2)}
\end{matrix}
\bigg) \!-\! \bigg(
\begin{matrix}
0.302\\
0.302
\end{matrix}
\bigg) \!\Bigg] \!= \!\Big(
\begin{matrix}
-0.012 \\
\;\;\,0.026
\end{matrix}
\Big)  \Big(\phi_{t-1}^{(1)} - \phi_{t-1}^{(2)} - 0.152 \Big) +  \; \mbox{higher order AR-terms} \; + \eta_{t}.
\end{equation}
In total the methods from cointegration work well with respect to phase synchronization on the dataset. One critical issue is that the phases differences in the lower plot of Figure~\ref{fig:ElectrochemicalOscillator1} can hardly be viewed as Gaussian random variables. A density plot shows clearly a skewed distribution for the phase differences which potentially leads to different p-values of the test statistics. To investigate this we have simulated the distribution of the error-correction coefficients $\alpha_i$ under the hypothesis $\alpha_i=0$ in this example with a parametric VAR-bootstrap. The results surprisingly indicate that the Gaussian approximation is reasonable. Details are omitted since this topic needs a deeper investigation which is postponed to future work.


\section{Conclusion and outlook}\label{sec:conclusion}

We have pointed out the connection between the theory of cointegration and the theory of phase synchronization. In particular, a cointegrated dynamical system can be used as a stochastic model for a multivariate phase process which describes the behavior of the phases in detail. Contrary to other concepts like the spectral coherence which only describe the properties of the phase processes, we now have a specific model in the time domain for the dynamics of the phase processes. The model leads to the characterization and identification of the equilibrium relations related to phase synchronization: For example, the unknown coupling structure can be revealed by using tests for unidirectional and bidirectional coupling. We have demonstrated this in applications for a coupled R\"{o}ssler-Lorenz system with noise and electrochemical oscillators.

In neuroscience where phase synchronization is regarded as essential for functional coupling of different brain areas the new methods coming from cointegration may enhance the present methods like spectral analysis, correlation and coherence analysis, triplet analysis, joint phase histograms etc. This however needs to be investigated and confirmed.

From a physical view the potential of cointegration for phase synchronization can perhaps best be understood from a comparison with the Kuramoto model. As pointed out in this paper, cointegration can be understood as an approximation to a stochastic Kuramoto-type model which can be fitted to the data (apart from the different treatment of phase shifts - see below). The cointegrated system describes both: 1) how the system is constantly pulled towards the equilibrium determined by the phase synchronization relations; and 2) how the system is constantly disturbed by little shocks and kicked out out of the equilibrium.

Another important aspect is that the cointegration model covers both: the case where the phases are synchronized and the case where they are not synchronized. As a consequence, the diagnosis on synchronization can be made from a statistical fit of the model to the data.

At the same time, the comparison to the Kuramoto model shows the difference to previous research: The Kuramoto model (\ref{stochasticKuramoto}) has the equilibrium $\phi^{(j)} - \phi^{(i)} \equiv 0 \mbox{ mod } 2\pi \;\forall \,i,j$ while the cointegration model (\ref{KuramotoCointegration}) has the equilibrium $\phi^{(j)} - \phi^{(i)} = 0 \;\forall \,i,j$. This means two processes whose phase difference is shifted (quickly) by $2\pi$ may be diagnosed as phase synchronized in the Kuramoto model while they will not be diagnosed as phase synchronized in the cointegration model. One may term this strong/weak synchronization. We advocate the view that phase jumps as in the R\"{o}ssler-Lorenz system
hardly occur with real data meaning that the cointegration model is a proper model in most cases. Nevertheless, it would be important to have similar techniques and results for the stochastic Kuramoto model. Another aspect is that phase jumps may occur erroneously caused by noise or wrong data.
Of course this raises the question of robustness of the phase estimate.

There is a wealth of possible directions for future research: An important issue is a theoretical treatment of the present model taking into account that the phases are
unobserved and the present cointegration model describes the dynamics of the phases as the unobserved state in a nonlinear state space model (VEC-state oscillators - see
Section~\ref{sec:Appendix1}). It seems to be challenging to derive, e.g., the asymptotic distribution of the likelihood ratio test for phase synchronization
in such a setting. Another topic of interest is a larger number of oscillators. In some example (for example for phase synchronization of the cardiovascular and respiratory system)
the correcting forces are obviously different if the system is out of the equilibrium to the positive or negative side. Here a nonlinear cointegration
system would be needed for a proper modeling.

\section{Appendix 1: VEC - state oscillators}\label{sec:Appendix1}

Inspired by the motivation at the end of the introduction we define the class of \linebreak \textit{VEC - state oscillators} as a general model for (possibly phase synchronized) d-dimensional oscillators with random phases. The model is the nonlinear state space model with

\medskip

\noindent \underline{State Equation:}
\begin{equation} \label{StateEQ}
\Delta \phi_t = \Pi \, \phi_{t-1} + \sum_{i=1}^{p-1} \Gamma_i \,
\Delta \phi_{t-i} + \mu + \eta_t, \quad \eta_t \sim \mathcal{N}(0,\Sigma_\eta),
\end{equation}

\noindent (or, alternatively, a model which guarantees positivity of the phases - see below)
\medskip

\noindent \underline{Observation Equation:}
\begin{equation} \label{ObesrvationEQ}
Y_t^{(i)} = a_i \cos(\phi_t^{(i)}) + b_i + \varepsilon_t^{(i)}, \quad \varepsilon_t \sim \mathcal{N}(0,\Sigma_\varepsilon)
\end{equation}

\noindent where $\varepsilon_t$ and $\eta_t$ are mutually and serially independent. In its simplest form the amplitude $a$ and the baseline $b$ are parameter vectors. In some cases they need to be time varying (see (ii) below). The case ${\rm rank}(\Pi)=0$ is included (the case ${\rm rank}(\Pi)=d$ requires the additional term $\nu t$ in (\ref{StateEQ}) for a meaningful phase-model - see Section~\ref{sec:newdefinition}). If $1 \le {\rm rank}(\Pi)\le d-1$ we have phase synchronization. In that case we prefer writing the state equation in the form
\begin{equation} \label{VecStateEQ}
(\Delta \phi_t - \omega_0) = \alpha \, \big(\beta'  \phi_{t-1} - \beta_0\big) + \sum_{i=1}^{p-1} \Gamma_i \,
\big(\Delta \phi_{t-i} - \omega_0 \big) + \eta_{t}
\end{equation}
(see Section~\ref{sec:cointegration}) with the phase synchronization relations $\beta'  \phi_{t} - \beta_0 =0$, the mean phase shift $\beta_0$, and the deterministic trend $\omega_0 t$. $\Pi$ is now decomposed into $\Pi = \alpha \beta'$ with $d \times r$ - matrices $\alpha$ and $\beta$ of full rank $r$.

The most common approach in physics is to estimate the phase process $\phi_t^{(i)}$ from the oscillator $Y_t^{(i)}$ by means of the Hilbert transform. In terms of state space models one may regard this as a smoother for estimating the mean of the conditional distribution (although its  properties are not clear). The approach of this paper is to use the estimated phases ``as if they were observed'' and to apply standard cointegration techniques to it.

Of course the effect of the Hilbert transform on the properties of these estimates is not clear, and the significance levels of the tests need to be investigated. This is beyond the scope of this paper, but we want to provide some heuristic arguments why everything remains the same even in this situation: Taking the conditional expectation (given $Y_{1},\ldots,Y_{n}$) in (\ref{VecStateEQ}) reveals that the linear structure of the conditional means is the same as before. The difference is that the variance of the innovations will be smaller (the estimated phases are smoother than the original ones) and the errors become dependent. A least squares regression based on the conditional expectations should also lead to consistent estimates for the parameters and in particular the t-statistic for a specific parameter (which is rescaled by the empirical variance) should follow asymptotically a t-distribution. The argument is the same for the ADF-test in the case $a \neq 0, \, b=0$ since this is the ``regular'' case where the drift dominates the stochastic trend and the asymptotic distribution of the unit root estimate is Gaussian (\cite{Ful-1996}, Theorem 10.1.5).

For the LR-test with its nonstandard distribution a different heuristics is needed: Inspection of the proof of the asymptotic distribution of the LR-test (\cite{Joh-1995}, Theorem 11.1) reveals that its limiting distribution only depends on the number $d\!-\!r$ of common stochastic trends and on the model for the deterministic terms. The model for the deterministic terms is fixed - so we only have to check that the number of stochastic trends remains the same for the estimated phases (estimated with the Hilbert transform) as for the original unobserved phases. To check this we consider in this heuristics only the case of $1\!:\!1$ phase synchronization relations, i.e. where the system splits up into $d\!-\!r$ oscillators and each of the remaining $d$ oscillators is stuck  to one of the former ones by the cointegration relation, i.e. we have $d\!-\!r$ groups of oscillators of different size (this is a common case for phase synchronization). We now have only to check that the groups of phase processes ``stay together'' under the Hilbert transform. First it is obvious that an integrated process stays integrated: The phases will constantly increase and a trend stationary process can be excluded because this is the case where the length of the cycle is almost stuck to a pregiven length and it is clear that a true phase process with variable cycles will not be transformed by the Hilbert transform into a phase process with almost constant cycle-length. In addition it is also clear that the phase processes will stay in the same group, i.e. they will not jump to another group with a different trend since this is clearly reflected by the number of cycles. Thus the number of stochastic trend should stay the same under the Hilbert transform.

We emphasize that these are only heuristic arguments. A mathematical proof that the asymptotic distributions of the test statistics stay the same would be highly welcome. It is obvious that deriving such a result may be very challenging.

A more sophisticated approach seems to be to estimate the phases in the above model by means of a particle filter. This has been done in \cite{Dah-2017} in the univariate case $d=1$ (in the more general setting (i)-(iii) from below). However, also with this approach the situation is not clear: A proper test for synchronization would then e.g. be a likelihood ratio test based on the original observations $Y_t$. The test statistic for such a test could be approximated by means of a particle filter - but its distribution under the null hypothesis is also not clear and difficult to derive.

Our personal view is that for systems with stronger noise the above particle filter or a periodogram based method lead to better estimation results. For chaotic oscillators as in
Section~\ref{sec:application} the Hilbert transform seems to be the better choice. Furthermore the Hilbert transform is used in most applications.

\subsection*{Generalizations of the state space model}

\noindent \underline{(i) Other oscillation patterns:}

\noindent For non-cosine type signals one may use instead the observation equation
\begin{equation*}\label{}
Y_t = a_t \, f(\phi_t) + b_t + \varepsilon_t, \qquad \quad
\end{equation*}
with $f$ being a $2 \pi$-periodic real valued function representing the oscillation pattern (cf. \cite{Dah-2017}). $f$ typically is unknown and needs to be estimated. An example are ECG-data. In addition one may need a time varying amplitude and a baseline. This can be achieved with a state space model also for $a_t$ and $b_t$, for example a VAR-model with mean different from zero.

\medskip

\noindent \underline{(ii) Positivity of the phase increments:}

\noindent In nearly all cases it is clear that the phase increments of an oscillator should be positive. However, in the above model with Gaussian noise also negative phase increments may occur. This happens with small probability since $\omega_0$ usually is large in comparison with the standard deviation of the $\eta_t$. Furthermore, this should not be a big problem if the model is mainly used for estimation and testing (and not for simulation).

To overcome this problem we have used in \cite{Dah-2017} in the case of a single oscillator an integrated ACD (autoregressive conditional duration) model for the state. At present we are having no idea how this ACD-model can be extended to the cointegrated case (see however \cite{Mos-2005} for bivariate ACD-models).

\section{Appendix 2: Computational and modeling aspects}\label{sec:Appendix2}

\subsection {Computational aspects} \label{sec:Appendix2:computational}

\noindent There exists several software programmes on cointegration analysis. For example the CATS module within the RATS Econometrics Software (http://www.estima.com/) and OxMetrics (http://www.oxmetrics.net/). For the analysis of the R\"{o}ssler-Lorenz system in Section~\ref{sec:application} we have used the R-packages \textit{urca} \; (http://cran.r-project.org/web/packages/urca/index.html - see \cite{Pfa-2008}) and \textit{systemfit} \; (http://cran.r-project.org/web/packages/systemfit/index.html - see \cite{Hen-2007}). Here are some details:

\bigskip

\baselineskip1.4em
\noindent \underline{1. Initial situation:}
The unwrapped phases are given as vectors phi1, phi2 (e.g. determined by the Hilbert-transform). The first values phi1[1] and phi2[2] should both lie in $[0,2\pi)$.\\[6pt]
\underline{2. Johansen LR-test (top-bottom version):}\\
require(urca)\\
joha.test <- ca.jo(cbind(phi1,phi2),type = "trace", ecdet = "none", K = 2, spec="transitory")\\
print(summary(joha.test))\\[6pt]
\underline{Note:} The specification \textit{ecdet = "none"} fits (\ref{Pi-representation}) with $\mu$ i.e. with $\omega_0$ and $\beta_0$ (Case 3 in \cite{Jus-2006}, Chapter 6.3) (while \textit{ecdet = "const"} and \textit{ecdet = "trend"} correspond to case 2 and case 4, respectively). We recommend using the quantiles from the appendix of \cite{Jus-2006} (Case 3) instead of the printed values in the summary.\\[6pt]
\underline{2a. ADF-test for testing phase synchronization}\\\hspace*{0.45cm}\underline{(testing the phase-difference for \textit{integrated process}  vs. \textit{stationary process}):}\\
require(urca)\\
beta2 <- 1 \quad \# Specify cointegration vector beta\\
adf.test <- ur.df(phi1 - beta2 * phi2, type="drift")\\
print(summary(adf.test))\\[6pt]
\underline{Note:} The specification \textit{type="drift"}  corresponds to Case 3 in \cite{Ham-1994}, p. 529. The relevant statistics is tau2 which should be compared with the quantiles of the standard normal distribution (for sample sizes less than 300 the t-distribution probably gives the better quantiles). The quantiles printed in the summary correspond to Case 2 in \cite{Ham-1994} where a different hypothesis is tested with the same statistics.\\[6pt]
\underline{3. and 3a. Estimation, further testing, and unidirectional coupling:}\\
require(systemfit)\\
T <- length(phi1)\\
err.lag1 <- (phi1-beta2*phi2)[-c(T-1, T)]\\
err.lag2 <- (phi1-beta2*phi2)[-c(T-1, T)]\\
dphi1 <- diff(phi1)\\
dphi2 <- diff(phi2)\\
diff.dat <- data.frame(embed(cbind(dphi1, dphi2), 2))\\
colnames(diff.dat) <- c('dphi1', 'dphi2', 'dphi1.1', 'dphi2.1')\\
eqPhi1 <- dphi1 $\sim$ err.lag1 + dphi1.1 + dphi2.1\\
eqPhi2 <- dphi2 $\sim$ err.lag2 + dphi2.1 + dphi1.1\\
system <- list(phi1 = eqPhi1, phi2 = eqPhi2)\\
estSystem <- systemfit(system, data=diff.dat)\\
print(summary(estSystem))\\[6pt]
\underline{4. Final estimation of the reduced system:}\\
eqPhi1 <- dphi1 $\sim$ dphi1.1\\
eqPhi2 <- dphi2 $\sim$ err.lag2 + dphi2.1\\
system <- list(phi1 = eqPhi1, phi2 = eqPhi2)\\
estReducedSystem <- systemfit(system, data=diff.dat)\\
print(summary(estReducedSystem))


\subsection {Modeling aspects} \label{sec:Appendix2:modeling}

One needs to be aware of the fact that the VEC-model has been applied in Section~\ref{sec:application} in the highly misspecified situation of two noisy chaotic oscillators. Our goal was to demonstrate that already a ``simple'' VEC-model can be used to successfully identify phase synchronization  and unidirectional coupling. For that reason we have chosen the fixed order $p=2$. In Figure~\ref{fig:RoesslerLorenzPhases} the oscillation of the phase
of the Lorenz attractor is due to the slightly steeper ascent
than the descent of the Lorenz attractor. In an additional data analysis (not reported here) we have taken this into account by
choosing the higher model order $p=4$ for the AR-part. This has led to a clear improvement of the fit
for the Lorenz attractor while higher model orders than $p = 4$ only showed smaller improvements.
We therefore have redone the complete analysis of Section~\ref{sec:application} with $p = 4$. There was
a clear quantitative improvement in the model fit but qualitatively all results stayed the same.

In Section~\ref{sec:application2} we have chosen the higher order $p = 6$ due to the in-cycle
fluctuation of the phases and the phase-differences. Here the model order was chosen by
inspection of the significant lags in the parametric model fit.

A common suggestion in time series analysis is to use a model selection criterion for choosing $p$ such
as the AIC or the BIC (cf. \cite{Gre-2008}, p.752). In the case of the nonlinear R\"{o}ssler-Lorenz system
investigated in Section~\ref{sec:application} we have found with the BIC the exorbitant
model order $p=31$ and with the AIC even $p=95$ for the phases of the Lorenz attractor (although not
all lags were significant). This reflects the fact that we need a very high order to obtain a good approximation
of the misspecified system with a VEC-model. On the other hand we would hardly trust the outcome of
significance tests etc. with such a large order. For that reason one has to refrain from these automatic
procedures for chaotic oscillators and look for a more parsimonious model order. Of course this topic
requires deeper investigations.


\bigskip
\bigskip

\noindent \textbf{Acknowledgements}\\[6pt]
\noindent {\small This work was part of the project ``Econometric Models for Phase Synchronization'' supported by the University of Heidelberg under the grant \textit{Frontier} D.801000/08.023 submitted in March 2008.\\[4pt]
\indent {\small The first author had the privilege to be from 2001--2007 a member of the DFG-priority programme SPP 1114 ``Mathematical methods for time series analysis and digital image analysis'' (jointly with many physicists). In this joint project he learned about phase synchronization from several colleagues including J\"{u}rgen Kurths, Ulrich Parlitz, Bj\"{o}rn Schelter and Jens Timmer - see \cite{Dah-2008}.}\\[4pt]
\indent {\small We are very grateful to S{\o}ren Johansen for some comments on an earlier version of this paper and to Florian M\"{u}ller-Dahlhaus for extensive discussions on the role of phase synchronization in neuroscience.}

\medskip

\noindent  \textbf{Disclaimer:} The views expressed here are those of the authors and not
necessarily those of its employers.

\bibliographystyle{apalike}	
\bibliography{phaseSynchronization111arXiv}
\end{document}